\documentclass{ejm}

\usepackage{graphicx}
\usepackage{epsfig}
\usepackage{epstopdf}
\usepackage{amsmath}
\usepackage{amssymb}
\usepackage{hyperref}
\usepackage{upmath}

\usepackage{rcs}
\RCS $RCSfile: paper_gwi.tex,v $
\RCS $Revision: 1.5 $
\RCS $Date: 2012/12/09 07:24:26 $

\newcommand{\leavethisout}[1]{}

\title[Three-phase free boundary problem with melting and dissolution]{A
  three-phase free boundary problem with melting ice and dissolving gas}  

\author[M. Ceseri \and J. M. Stockie]{Maurizio Ceseri
  \ \and
  John M. Stockie
} 
\affiliation{
  Department of Mathematics, Simon Fraser University, 8888
  University Drive, Burnaby, British~Columbia, Canada,~V5A~1S6\\ 
  email\textup{\nocorr: \texttt{mceseri@sfu.ca}},
  \textup{\texttt{jstockie@sfu.ca}}} 

\begin{document}

\maketitle
\renewcommand{\thefootnote}{\fnsymbol{footnote}}


\newcommand{\units}[1]{\mbox{$\mathrm{#1}$}}
\newcommand{\bunits}[1]{[\units{#1}]}
\newcommand{\Stefan}{\texttt{St}}
\newcommand{\Lewis}{\texttt{Le}}
\newcommand{\Biot}{\texttt{Bi}}
  \renewcommand{\Stefan}{\operatorname{St}}
  \renewcommand{\Lewis}{\operatorname{Le}}
  \renewcommand{\Biot}{\operatorname{Bi}}
\newdefinition{remark}{Remark}
\newcommand{\chtc}{\theta} 
\newcommand{\meshvel}{u}     
\newcommand{\degC}{\units{{}^\circ\hspace*{-0.13em}C}}
\newcommand{\degK}{\units{{}^\circ\hspace*{-0.13em}K}}

\newcommand{\ds}[1]{{\displaystyle #1}}
\newcommand{\Cinit}{C_0}
\newcommand{\Cinner}{\gamma}
\newcommand{\Couter}{G}

\newcommand{\tabrange}[2]{[#1,#2]}
\renewcommand{\tabrange}[2]{}

\footnotetext[0]{Latest Revision: \RCSRevision\ (\RCSDate).  Printed: \today.}

\begin{abstract}
  We develop a mathematical model for a three-phase free boundary
  problem in one dimension that involves the interactions between gas,
  water and ice.  The dynamics are driven by melting of the ice layer,
  while the pressurized gas also dissolves within the meltwater.  The
  model incorporates a Stefan condition at the water-ice interface along
  with Henry's law for dissolution of gas at the gas-water interface.
  We employ a quasi-steady approximation for the phase temperatures and
  then derive a series solution for the interface positions.  A
  non-standard feature of the model is an integral free boundary
  condition that arises from mass conservation owing to changes in gas
  density at the gas-water interface, which makes the problem
  non-self-adjoint.  We derive a two-scale asymptotic series solution
  for the dissolved gas concentration, which because of the
  non-self-adjointness gives rise to a Fourier series expansion in
  eigenfunctions that do not satisfy the usual orthogonality
  conditions.  Numerical simulations of the original governing equations
  are used to validate the series approximations.
\end{abstract}


\section{Introduction}
\label{sec:intro}

This paper is concerned with a three-phase free boundary problem
involving interactions between ice, liquid water, and air.  The
water-ice interface is driven by a melting process, while the gas-water
interface is governed by dissolution of gas within the water phase.  The
primary phenomenon we are interested in capturing is the compression or
expansion of gas that occurs in response to the motion of phase
interfaces.
  
Free boundaries arise naturally in the study of phase change problems
and have been the subject of extensive study in the applied mathematics
literature~\cite{Crank1984,friedman-1982,friedman-2000,gupta-2003}.
Mathematical models of free boundaries generally involve solving partial
differential equations on some region(s), along with given boundary
conditions on a portion of the boundary; however, part of the domain
boundary remains unknown, and thus some additional relationship must
be provided to determine the free boundary.  A classical example is the
Stefan problem for a solid-liquid interface~\cite{Crank1984} that
describes a melting or solidification process.  Here, the primary
variable (temperature) is governed by the diffusion equation, while
the speed of the solid-liquid interface is related to the difference in
heat flux on either side, which is a statement of conservation of
energy.  Friedman~\cite{Friedman1959} established well-posedness and
regularity results for this melting problem, while Crank~\cite{Crank1984}
and Carslaw and Jaeger~\cite{Carslaw1988} derived analytical solutions
using Neumann's method for a variety of physical applications.  A
characteristic feature of all of these solutions is that the speed of
the free boundary between the phases is proportional to
${t}^{1/2}$, where $t$ is elapsed time.

Another class of free boundary problems occurs in the study of
dissolution and cavitation of gas bubbles immersed in
fluid~\cite{plesset-prosperetti-1977}.  Friedman~\cite{Friedman1960}
studied the interface evolution for a spherically-symmetric gas bubble
immersed in a water-filled container of infinite extent, and he proved
existence, uniqueness and regularity of the solution.
Keller~\cite{Keller1964} studied a similar problem and determined the
conditions under which multiple gas bubbles are stable.  In particular,
he found that gas bubbles should either collapse or else grow
indefinitely in an infinite medium.  In a closed container, however,
bubbles can reach a stable equilibrium state and he proved that the only
stable solution is the one with a single bubble.

\leavethisout{
[[ IS THIS PARAGRAPH REALLY NEEDED? ]] Since the gas bubble can be
compressed during phase change, the density jump across the water-ice
interface has to be taken into account.  The particular feature of a
multiphase system with phases of different densities is the relative
movement of phases that come into play to ensure mass conservation at
any interface.  In the report \cite{Wilson1982}, the author defines a
coordinate system which is at rest between the different phases.  Using
a different coordinate transformation, Tao \cite{Tao1979} recast the
original two phase Stefan problem in a classical Stefan problem with
phases with the same density.
}

This paper was originally motivated by a recent modelling study of sap
exudation in sugar maple trees during the spring
thaw~\cite{Ceseri_Stockie2012}.  This is the process whereby maple (and
other related tree species) generate positive stem pressure that can
cause sap to seep out of any hole bored in the tree trunk.  In late
winter there are no leaves to drive transpiration, and the maple tree's
internal pressure generation mechanism is believed to derive from
thawing of frozen sap within libriform fiber cells located in the
sapwood or xylem~\cite{Milburn1984}.  These fibers are typically filled
with gas during the growing season, but during the onset of winter, ice
is believed to form on the inner fiber walls thereby compressing the gas
trapped within.  During the spring thaw, the ice layer melts thereby
freeing the compressed gas which is then free to re-pressurise the xylem
sap.  A mathematical model for sap exudation has recently been developed
in the paper~\cite{Ceseri_Stockie2012}, which contains more details
about the physical processes involved.  The model predicts build-up of
stem pressures sufficient to dissolve gas bubbles in the xylem sap,
which may also be related to the phenomenon of winter embolism recovery
that occurs in a much wider range of tree
species~\cite{konrad-rothnebelsick-2003,tyree-sperry-1989}.

In this paper, we consider a mathematical model for a simpler situation
in which a closed container is divided into three compartments
containing gas, water and ice, in that order.  While this scenario is
not identical to that seen in maple xylem cells, it is nonetheless close
enough that it permits us to study in detail the dynamics of the free
boundaries.  To our knowledge there has been no other similar study of
three-phase flow involving gas dissolution and ice melting.  There are
several other problems arising in porous media flow that have some of
the same features as our model.  For example, the modeling of marine gas
hydrates~\cite{tsypkin2000,xu-2004} involves the interplay between
gaseous and solid hydrates, water, and possibly other components flowing
within porous sediments.  Although these models involve a Stefan
condition for a melting front, the gas dynamics are driven by hydrate
dissociation instead of gas dissolution.  Another related problem arises
in the freezing and thawing of soils contaminated by non-aqueous phase
liquids (or NAPLs)~\cite{Huyakorn1994,Singh2012}.  Here, there is a
dissolved gas component but the problem is complicated further by the
presence of additional phases as well as effects such as mixed
wettability.

The purpose of the present work is to analyze a simple three-phase model
that incorporates the dynamics of melting and dissolution.  The model is
introduced in Section~\ref{sec:model} and reduced to non-dimensional
form.  A numerical algorithm is described in
Section~\ref{sec:numerics}, and simulations in Section~\ref{sec:results}
yield insight into the behaviour of the solution.  Motivated by these
results, we then derive an asymptotic solution in
Section~\ref{sec:asymptotics} that captures the essential dynamics, and
comparisons are drawn with the full numerical solution.  Our main aim
in this work is three-fold:
\begin{itemize}
\item To understand the basic phase interface dynamics and identify the
  relevant dimensionless quantities and time scales;
\item To develop approximate analytical solutions that can be used
  either to design more efficient numerical schemes or to up-scale
  material coefficients for microscale models such
  as~\cite{Ceseri_Stockie2012};
\item To draw connections with existing results on bubble dissolution
  dynamics.
\end{itemize}

\section{Mathematical Model}
\label{sec:model}

\begin{figure}
  \begin{center}
    \includegraphics[width=0.8\textwidth]{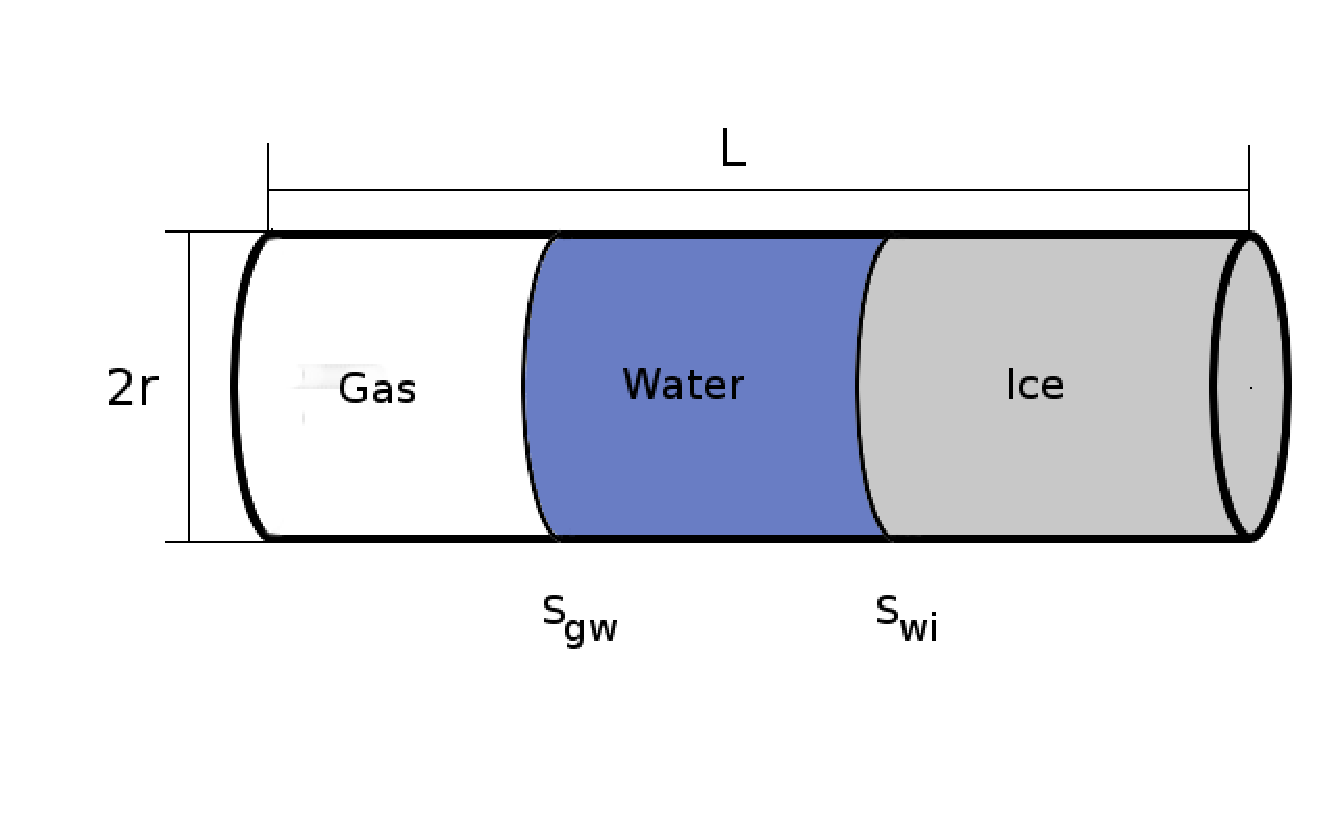}
  \end{center}
  \caption{Diagram depicting the cylindrical geometry and moving phase 
    interfaces.}
  \label{fig1}
\end{figure}

Consider a cylindrical container of constant radius $r$ and length $L$
(both measured in \units{m}) that is separated into three compartments
containing gas, water and ice as pictured in Figure~\ref{fig1}.  Assume
that the cylinder is long and thin so that $L\gg r$ and we can restrict
ourselves to a one-dimensional setting where the axial coordinate $x$
varies from $0$ to $L$.  There are two moving interfaces at locations
$x=s_{gw}(t)$ and $s_{wi}(t)$ that separate gas from water and water
from ice respectively.

For the sake of simplicity, we consider melting that is driven by a heat
source applied on the left-hand boundary; and although we will not
consider the freezing process, our model can be easily extended to
handle the freezing case.  We are thus interested in the following
physical phenomena: (1) heat transfer occuring within and between the
three phases; (2) phase change at the water-ice interface as ice melts
to form liquid water; and (3) dissolution of pressurised of gas at the
gas-water interface, with subsequent diffusion of dissolved gas in the
water compartment.  The moving boundaries are driven by two different
mechanisms.  The water-ice interface is driven by phase change and the
speed of the interface is proportional to the difference between heat
flux from the adjacent compartments (which is the classical \emph{Stefan
  condition}~\cite{Crank1984}).  On the other hand, the gas-water
interface moves in response to changes in volume not only from the
dissolution of gas in water, but also from the volume change owing to
the density difference between water and ice.

We next list a number of simplifying assumptions:
\begin{enumerate}
\renewcommand{\theenumi}{A\arabic{enumi}}
\item The \label{assume:insulated} lateral surface of the cylindrical
  domain is thermally insulated so that heat flows only in the axial
  ($x$) direction.
\item The\label{assume:close} system is closed so that the total mass of
  gas (free plus dissolved) is constant.  The total mass of liquid and
  frozen water also remains constant, although the mass of the
  individual phases may change in time.
\item The water\label{assume:density} and ice densities are constant and
  are not affected by changes in temperature.
\item Diffusion \label{assume:gasdiff} in the gas compartment is fast
  enough in relation to other processes that the gas density can be
  taken as a function of time only. Indeed, considering the
  self-diffusion coefficient for air ($D=2\times 10^{-5}~\units{m^2/s}$)
  and a typical length scale ($d=100~\units{\mu m}$), the time scale for
  air diffusion is roughly $t\approx d^2/4D=1.25\times
  10^{-4}~\units{s}$ (see \cite[sect. 3.32]{Crank1956}).
\item The amount of gas\label{assume:icewater} that dissolves in the
  water compartment is small enough relative to the initial gas volume that
  gas dissolution does not significantly affect the motion of the
  gas-water interface.  When combined with the previous assumption, this
  implies that the motion of the gas-water interface is due only to the
  melting of ice and the subsequent density increase as ice changes
  phase from solid to liquid.
\item Neither gas nor water\label{assume:gasnoice} dissolve in or
  otherwise penetrate the ice layer.
\item Some water\label{assume:waterlayer} is always present in the
  fiber.  Instead of having initial conditions where all water is in the
  frozen state initially, a very thin layer of liquid is assumed to
  separate the gas from the ice.
\end{enumerate}

We note that many features of this problem are similar to the sap
exudation model derived in \cite{Ceseri_Stockie2012} for
a closed system that consists of two distinct classes of xylem cells:
libriform fibers, in which thawing of ice allows compressed gas to force
the melted sap through the porous fiber wall; and the neighbouring vessels
that contain gas and liquid sap, where the sap in turn contains both
dissolved gas and sucrose.  That sap exudation model differs from the
model we develop here in several respects:
\begin{itemize}
\item we treat only a single compartment containing three phases;
\item in \cite{Ceseri_Stockie2012}, the ice in the fiber is sandwiched
  between the gas and liquid compartments;
\item we do not consider osmotic effects or sap flow through the
  permeable fiber/vessel boundary; and
\item the simple cylindrical geometry permits us to neglect surface
  tension effects due to curvature of the gas-water interface.
\end{itemize}

In the next three sections, we derive the governing equations and
boundary conditions in each of the gas, liquid and ice compartments.
Following that, we summarise in a separate section the remaining
interfacial matching conditions that connect solutions on either
side of the moving boundaries. 

\subsection{Gas compartment}
\label{sec:gas}

Denote the temperature in the gas compartment by $T_g(x,t)$ \bunits{\degK} which
obeys the diffusion equation
\begin{subequations}\label{eq:Tgdim}
  \begin{gather}
    \rho_g c_g \frac{\partial T_g}{\partial t} = 
    \frac{\partial}{\partial x} \left(k_g\frac{\partial T_g}{\partial
        x}\right),   
    \label{eq:Tgdim-pde}
  \end{gather}
  for $0<x<s_{gw}(t)$ and $t>0$.  The constant parameters appearing in
  this equation are the thermal conductivity $k_g$ \bunits{W/m\,\degK} and
  specific heat $c_g$ \bunits{J/kg\, \degK}, while the gas density $\rho_g$
  \bunits{kg/m^3} depends on time according to
  Assumption~\ref{assume:gasdiff}.  The initial temperature distribution
  is given by
  \begin{gather}
    T_g(x,0)=T_{g0}(x), 
    \label{eq:Tgdim-ic}
  \end{gather}
  for $0<x<s_{gw}(0)$.  The heat source that drives the melting of the
  ice compartment is located at the left-hand boundary $x=0$ where we
  impose a constant temperature
  \begin{gather}
    T_g(0,t) = T_1 > T_c, 
    \label{eq:Tgdim-bc}
  \end{gather}
\end{subequations}
that is strictly greater than the melting temperature of ice,
$T_c=273.15\;\units{\degK}$.

Returning to the gas density, we will now use Assumptions
\ref{assume:close} and \ref{assume:gasdiff} to derive a closed-form
expression for $\rho_g(t)$ using a mass balance argument that considers
the total mass of gas (which must be constant) and its division between
the gas and water compartments.  First, the mass \bunits{kg} of
dissolved gas in the water compartment is
\begin{gather} 
  m_w(t) = A M_g \int_{s_{gw}(t)}^{s_{wi}(t)} C(x,t)\, dx,
  \label{eq:mw}
\end{gather}
where $M_g$ is the molar mass of air \bunits{kg/mol}, $A=\pi r^2$ is the
cross-sectional area of the cylindrical domain \bunits{m^2} and
$C(x,t)$ is the concentration of dissolved gas \bunits{mol/m^3} (whose
governing equation will be given in the next section).  The air density
may then be written as
\begin{align}
  \rho_g(t) = \frac{A \rho_g(0) s_{gw}(0) - m_w(t) + m_w(0)}{A
    s_{gw}(t)}, 
  \label{eq:density}
\end{align}
which along with \eqref{eq:mw} determines $\rho_g(t)$ once the dissolved
gas concentration and gas-water/water-ice interface positions are known.

\subsection{Water compartment}
\label{sec:water}

We next turn to the water compartment where the temperature $T_w(x,t)$ also
satisfies the heat equation
\begin{subequations}\label{eq:Twdim}
  \begin{gather}
    \rho_w c_w \left(\frac{\partial T_w}{\partial t}+v\frac{\partial
        T_w}{\partial x}\right) = \frac{\partial}{\partial
      x}\left(k_w\frac{\partial T_w}{\partial x}\right), 
    \label{eq:Twdim-pde}
  \end{gather}
  for $s_{gw}(t) < x < s_{wi}(t)$ and $t>0$, where $c_w$, $\rho_w$ and
  $k_w$ are the water specific heat, density and thermal conductivity
  respectively.  The extra heat convection term on the left hand side
  arises from the slow flow of water due to the melting of ice and the
  density difference between water and ice
  \cite{Crank1984,Tao1979,Wilson1982}; the convection velocity $v$
  \bunits{m/s} will be specified later in Section~\ref{sec:matching}.
  We also need to specify an initial temperature distribution
  \begin{gather}
    T_w(x,0) = T_{w0}(x).
    \label{eq:Twdim-ic}
  \end{gather}
\end{subequations}

The dissolved gas concentration $C(x,t)$ obeys the diffusion equation
\begin{subequations}\label{eq:Cdim}
  \begin{gather}
    \frac{\partial C}{\partial t} = \frac{\partial}{\partial x}
    \left(D_w\frac{\partial C}{\partial x}\right),
    \label{eq:Cdim-pde}
  \end{gather}
  where $D_w$ is the diffusion coefficient of air in water
  \bunits{m^2/s}.  At the gas-water interface, we impose Henry's law
  \begin{gather}
    C(s_{gw}(t),t) = \frac{H}{M_g}\rho_g(t),
    \label{eq:Cdim-henry}
  \end{gather}
  which states that the concentration of gas dissolved at the interface
  is proportional to the density of the gas in contact with the
  liquid.  Here, $H$ denotes the dimensionless Henry's constant.  
  Finally, we impose the initial condition
  \begin{gather}
    C(x,0) = \Cinit(x), 
    \label{eq:Cdim-ic}
  \end{gather}
  and the following no-flux boundary condition at the water-ice
  interface
  \begin{gather}
    \frac{\partial C}{\partial x}(s_{wi}(t),t) = 0,
    \label{eq:Cdim-bc}
  \end{gather}
  which is a simple statement of the fact that dissolved gas does not
  penetrate the ice (in accordance with Assumption
  \ref{assume:gasnoice}).
\end{subequations}

The following two remarks relate to the distribution of air between the
gaseous and dissolved phases.

\begin{remark}[Conservation of air]
  \itshape
  We first show that \eqref{eq:density} implies conservation of mass for
  total air in the gaseous and dissolved phases.  The total mass of air
  at any time $t$ can be written as the sum of the air in the water and
  gas compartments:
  \begin{align*}
    m &= m_w(t) + A \int_0^{s_{gw}(t)} \rho_g(t) \, dx, \\
    &= m_w(t) + A s_{gw}(t) \rho_g(t).\\
    \intertext{Then, replacing $\rho_g(t)$ with \eqref{eq:density} leads to}
    m &= m_w(0) + A \rho_g(0) s_{gw}(0).
  \end{align*}
  This last expression is simply the sum of the total initial mass of
  dissolved and gaseous air, and hence the total mass $m$ of air is
  conserved.
\end{remark}

\begin{remark}[Connection with Keller's analysis of gas bubble dynamics]
  \itshape Our aim here is to derive an expression for the rate of
  change of the gas density, which can then be related directly to an
  equation derived by Keller for the dynamics of dissolving gas bubbles
  in water \cite{Keller1964}.  To this end, we take the time derivative
  of the gas density from equation~\eqref{eq:density}
  \begin{align}
    \frac{d\rho_g}{dt} &= \frac{-\dot{m}_w(t) s_{gw}(t) -
      \dot{s}_{gw}(t) \big[ A \rho_g(0) s_{gw}(0) - m_w(t) +
        m_w(0) \big]}{A s_{gw}^2(t)},\nonumber\\
    &= - \frac{\dot{m}_w(t)}{A s_{gw}(t)}
    - \frac{\rho_g(t) \dot{s}_{gw}(t)}{s_{gw}(t)}, 
    \label{eq:drhogdt}
  \end{align}
  where the ``dot'' denotes the time derivative.  An
  expression for $\dot{m}_w(t)$ can be obtained by differentiating
  \eqref{eq:mw}
  \begin{gather*}
    \dot{m}_w(t) = A M_g \left[ \dot{s}_{wi}(t) C(s_{wi}(t),t) -
      \dot{s}_{gw}(t) C(s_{gw}(t),t) + \int_{s_{gw}(t)}^{s_{wi}(t)}
      \frac{\partial C}{\partial t}(x,t)\, dx \right].
  \end{gather*}
  The integral term containing $\partial_t C$ may be integrated
  directly by first replacing $\partial_t C$ using the concentration equation
  \eqref{eq:Cdim-pde}
  \begin{gather*}
    \dot{m}_w(t) = A M_g \Big[ \dot{s}_{wi}(t)
    C(s_{wi}(t),t) - \dot{s}_{gw}(t) C(s_{gw}(t),t) + D_w \Big(
    \frac{\partial C}{\partial x} (s_{wi}(t),t) -
    \frac{\partial C}{\partial x} (s_{gw}(t),t) \Big) \Big],
  \end{gather*}
  and then applying the boundary condition \eqref{eq:Cdim-bc}
  \begin{gather*}
    \dot{m}_w(t) = A M_g \Big[ \dot{s}_{wi}(t) C(s_{wi}(t),t) - \dot{s}_{gw}(t)
    C(s_{gw}(t),t) - D_w \frac{\partial C}{\partial x}(s_{gw}(t),t) \Big].
  \end{gather*}
  We may now substitute this last expression for $\dot{m}_w(t)$ into
  equation \eqref{eq:drhogdt} to obtain
  \begin{gather}
    \frac{1}{M_g}\frac{d(\rho_g s_{gw})}{dt} = 
    \dot{s}_{gw} C(s_{gw},t) - \dot{s}_{wi} C(s_{wi},t) + D_w
    \frac{\partial C}{\partial x}(s_{gw},t).
    \label{eq:simKeller}
  \end{gather}
  This equation coincides with Keller's
  equation~(2.6)~\cite{Keller1964}, except for slight differences
  arising from to the fact that we are working in a cylindrical geometry
  and we also include the water-ice interface motion in the evolution of
  the gas density.
\end{remark}

\subsection{Ice compartment}
\label{sec:ice}

In the ice compartment, the equation governing the ice temperature $T_i(x,t)$
is 
\begin{subequations}
  \begin{gather}
    \rho_i c_i\frac{\partial T_i}{\partial t} = 
    \frac{\partial}{\partial x}\left(k_i\frac{\partial T_i}{\partial
        x}\right),  
  \end{gather}
  for $s_{wi}(t)<x<L$ and $t>0$, where $c_i$, $\rho_i$ and $k_i$ are the
  specific heat, density and thermal conductivity of ice.  The initial
  temperature distribution is given 
  \begin{gather}
    T_i(x,0)=T_{i0}(x),
  \end{gather}
  and on the right boundary we impose a convective condition of
  the form
  \begin{gather} 
    -k_i \frac{\partial T_i}{\partial x}(L,t) = \chtc(T_i-T_2), \qquad
  \end{gather}  
  where $\chtc$ is a convective heat transfer coefficient \bunits{W/m^2\,
    \degK} and $T_2$ is a given ambient temperature.
\end{subequations}

\subsection{Interfacial and matching conditions}
\label{sec:matching}

We now state the matching conditions at the two phase interfaces.
First, we require that the temperature and heat flux are both continuous
at the gas-water interface
\begin{gather}
  T_g(s_{gw}(t),t) = T_w(s_{gw}(t),t),
  \label{eq:match-T-gw}
  \\
  k_g\frac{\partial T_g}{\partial x}(s_{gw}(t),t) =
  k_w\frac{\partial T_w}{\partial x}(s_{gw}(t),t) .
  \label{eq:match-dTdx-gw}
\end{gather}
Based on geometric and  conservation arguments, we can derive 
an equation for the evolution of the gas-water interface
\begin{gather}
  \dot{s}_{gw}(t)=\left(1-\frac{\rho_i}{\rho_w}\right)\dot{s}_{wi}(t), 
  \label{eq:sgw-dim}
\end{gather}
which relates the velocities of the two interfaces via the difference in
volume owing to contraction and expansion of ice (details of the
derivation are provided in Appendix~\ref{app:1}).  This condition is
also given in Crank's book~\cite[Eq.~1.32]{Crank1984} which expresses
the velocity of the liquid phase when a density change is taken into
account; therefore, we impose
\begin{gather}
  v=\dot{s}_{gw}
  \label{eq:v-sdot}
\end{gather}
for the convection speed in equation~\eqref{eq:Twdim-pde}.

At the water-ice interface, the temperature must be continuous
\begin{gather}
  T_w(s_{wi}(t),t) = T_i(s_{wi}(t),t) = T_c,
  \label{eq:match-T-wi}
\end{gather}  
with the added requirement that the temperature on both sides of the
interface must equal the melting point.  The evolution of the water-ice
interface is governed by 
\begin{gather}
  \lambda\rho_i \dot{s}_{wi} = k_i\frac{\partial T_i}{\partial x} -
  k_w\frac{\partial T_w}{\partial x}
  \qquad \text{at $x=s_{wi}(t)$},  
  \label{eq:stefan-dim}
\end{gather}
where $\lambda$ is the latent heat of melting per unit mass
\bunits{J/kg}.  This Stefan condition is a statement of conservation of
energy, where the amount of heat generated by the change of phase
($\lambda\rho_i\dot{s}_{wi}$) is balanced by the difference in heat
flux from either side of the phase interface.  Finally, to close the
system we require initial conditions for the phase interface locations:
\begin{gather}
  s_{gw}(0) = s_{gw0},\\
  s_{wi}(0) = s_{wi0}.
\end{gather}

\subsection{Parameter values}
\label{sec:params}

The values of all parameters defined above are given in
Table~\ref{tab:1} in SI units and are taken from the data for the sap
exudation model in \cite{Ceseri_Stockie2012}.  The geometrical
parameters $L=10^{-3}\;\units{m}$, $r=3.5\times 10^{-6}\;\units{m}$ and
$A=\pi r^2=3.85\times 10^{-11}\;\units{m^2}$ are all based on the size of a
typical libriform fiber in the xylem of a maple tree.

\begin{table}
  \caption{Physical parameters, with numerical values taken from
    \cite{Ceseri_Stockie2012}.} 
  \label{tab:1}
  \begin{minipage}{\textwidth}
    \tabcolsep=8pt
    \begin{tabular}{lccc}
      \hline\hline
      Parameter & Symbol & Units & Value \\
      \hline
      Domain length          & $L$          & \units{m}        & $1.0\times 10^{-3}$\\
      Domain radius          & $r$          & \units{m}        & $3.5\times 10{-6}$\\ 
      Cross-sectional area   & $A=\pi r^2$  & \units{m^2}      & $3.85\times 10^{-11}$ \\
      Densities              & $\bar{\rho}_g$ & \units{kg/m^3} & $1.29$ \\
                             & $\rho_w$     & \units{kg/m^3}   & $1000$ \\
                             & $\rho_i$     & \units{kg/m^3}   & $916$ \\
      Specific heats         & $c_g$        & \units{J/kg\, \degK} & $1005$ \\
                             & $c_w$        & \units{J/kg\, \degK} & $4180$ \\
                             & $c_i$        & \units{J/kg\, \degK} & $2050$ \\
      Thermal conductivities & $k_g$        & \units{W/m\,\degK}   & $0.0243$\\
                             & $k_w$        & \units{W/m\,\degK}   & $0.58$ \\
                             & $k_i$        & \units{W/m\,\degK}   & $2.22$ \\
      Thermal diffusivities, $\alpha = k/(\rho c)$ 
                             & $\alpha_g$   & \units{m^2/s}    & $1.87\times 10^{-5}$ \\
                             & $\alpha_w$   & \units{m^2/s}    & $1.39\times 10^{-7}$ \\
                             & $\alpha_i$   & \units{m^2/s}    & $1.18\times 10^{-6}$ \\
      Diffusivity of dissolved air in water & $D_w$   & \units{m^2/s}    & $2.22 \times 10^{-9}$\\
      Molar mass of air      & $M_g$        & \units{kg/mol}   & $0.0290$ \\
      Henry's constant       & $H$          &                  & $0.0274$  \\
      Convective heat transfer coefficient & $\chtc$ & \units{W/m^2\, \degK} & $10.0$ \\
      Latent heat of melting & $\lambda$    & \units{J/kg}     & $3.34\times 10^5$ \\
      Critical (melting) temperature of ice & $T_c$ & \units{\degK}  & $273.15$ \\
      Left boundary temperature  & $T_1$    & \units{\degK}    & $T_c+0.005$\\
      Right boundary temperature & $T_2$    & \units{\degK}    & $\tabrange{T_c-0.02}{T_c-0.005}$\\
      \hline\hline
    \end{tabular}
    \end{minipage}
\end{table}

\subsection{Non-dimensional equations}
\label{sec:nondim}

We now non-dimensionalise the governing equations by introducing the
following dimensionless quantities denoted by a superscript asterisk
($*$): 
\begin{gather}
  \begin{array}{cccc}
    x=Lx^*,         & s_{gw}=Ls^*_{gw}, & & s_{wi}=Ls^*_{wi}, \\
    t=\bar{t}\,t^*, & C=\bar{C}C^*,     & & \rho_g = \bar{\rho}_g \rho_g^*,\\
    T_\ell=T_c+(T_1-T_c)T_\ell^*.
  \end{array}
  \label{eq:nondim}
\end{gather}
We choose as a length scale $L=10^{-3}\;\units{m}$, which corresponds to
the typical length of a libriform fiber \cite{Ceseri_Stockie2012}.
Density is rescaled by the value $\bar{\rho}_g=1.29\;\units{kg/m^3}$ for
air at $1\;\units{atm}$ and $0\degC$, and the concentration by
$\bar{C}=\bar{\rho}_g/M_g$.  The time scale $\bar{t}$ is chosen equal to
the characteristic scale typical in Stefan problems for motion of the
water-ice interface 
\begin{gather}
  \bar{t} = \frac{L^2 \lambda \rho_i}{k_w(T_1-T_c)}, 
  \label{eq:tbar}
\end{gather}
because the melting process is the driving mechanism for this problem.

Substituting the expressions from \eqref{eq:nondim} into the model
equations and dropping asterisks to simplify notation, we
obtain the following dimensionless system where all new 
parameters are listed in Table~\ref{tab:2}.

\begin{table}
  \centering
  \caption{Characteristic scales and dimensionless parameters.} 
  \label{tab:2}
  \begin{minipage}{\textwidth}
    \tabcolsep=8pt
    \begin{tabular}{cccc}
      \hline\hline
      Parameter & Expression & Units & Value \\
      \hline
      $\bar{\rho}_g$&                                     & \units{kg/m^3}  & 1.29 \\[7pt]
      $\bar{C}$     & $\ds{\frac{\bar{\rho}_g}{M_g}}$     & \units{mol/m^3} & 44.6 \\[7pt]
      $t_w$         & $\ds{\frac{L^2}{\alpha_w}}$         & \units{s}       & $7.21$ \\[7pt]
      $t_g$         & $\ds{\frac{L^2}{\alpha_g}}$         & \units{s}       & $5.35\times 10^{-2}$ \\[7pt]
      $t_i$         & $\ds{\frac{L^2}{\alpha_i}}$         & \units{s}       & $8.46\times 10^{-1}$ \\[7pt]
      $\bar{t}$     & $\ds{\frac{L^2 \lambda \rho_i}{k_w(T_1-T_c)} =
                            \frac{t_w\delta}{\Stefan}}$   & \units{s}       & $1.06\times 10^5$ \\[7pt]
      \hline
      $\delta$      & $\ds{\frac{\rho_i}{\rho_w}}$        & & 0.916 \\[7pt]
      $\eta$        & $\ds{\frac{k_w}{k_g}}$              & & 23.9 \\[7pt]
      $\psi$        & $\ds{\frac{k_i}{k_w}}$              & & 3.83 \\
      $\Biot$       & $\ds{\frac{L\chtc}{k_i}}$           & & $4.50\times 10^{-3}$ \\[7pt]
      $\Lewis$      & $\ds{\frac{\alpha_w}{D_w}}$         & & 62.5 \\[7pt]
      $\Stefan$     & $\ds{\frac{(T_1-T_c)c_w}{\lambda}}$ & & $6.26\times 10^{-5}$ \\[7pt]
      $\widetilde{T}_2$ & $\ds{\frac{T_2-T_c}{T_1-T_c}}$  & & $\tabrange{-4}{-1}$ \\[7pt]
      $\beta_g$     & $\ds{\frac{\alpha_g\bar{t}}{L^2}=\frac{\alpha_g\beta_w}{\alpha_w}}$ & & $1.97\times 10^6$\\[7pt]
      $\beta_w$     & $\ds{\frac{\alpha_w\bar{t}}{L^2}=\frac{\delta}{\Stefan}}$           & & $1.46\times 10^4$\\[7pt]
      $\beta_i$     & $\ds{\frac{\alpha_i\bar{t}}{L^2}=\frac{\alpha_i\beta_w}{\alpha_w}}$ & & $1.25\times 10^5$\\[7pt]
      \hline\hline
    \end{tabular}
    \end{minipage}
\end{table}

In the gas compartment, $0<x<s_{gw}(t)$:
\begin{subequations}\label{eq:Tgnondim}
  \begin{gather}
    \frac{\partial T_g}{\partial t} = 
    \beta_g \, \frac{\partial^2 T_g}{\partial x^2} , 
    \label{eq:Tgnondim-pde} 
  \end{gather}
  \begin{gather}
    T_g(x,0) = T_{g0}(x) \quad \mbox{for } 0<x<s_{gw}(0),
    \label{eq:Tgnondim-ic} 
  \end{gather}
  \begin{gather}
    T_g(0,t) = 1,
    \label{eq:Tgnondim-bc} 
  \end{gather}
\end{subequations}
where the dimensionless diffusion coefficient $\beta_g = \alpha_g
\bar{t}/L^2$ and $\alpha_g=k_g/(\bar{\rho}_g c_g)$ is the thermal
diffusivity of air.

In the water compartment, $s_{gw}(t)<x<s_{wi}(t)$, we have equations for both
temperature and concentration
\begin{subequations}\label{eq:Twnondim}
  \begin{gather}
    \frac{\partial T_w}{\partial t} + 
    \dot{s}_{gw} \frac{\partial T_w}{\partial x} =
    \beta_w \, \frac{\partial^2 T_w}{\partial x^2},
    \label{eq:Twnondim-pde} 
  \end{gather}
  \begin{gather}
    T_w(x,0) = T_{w0}(x) \quad \mbox{for } s_{gw}(0)<x<s_{wi}(0),
    \label{eq:Twnondim-ic}
  \end{gather}
\end{subequations}
\begin{subequations}\label{eq:Cnondim}
  \begin{gather}
    \frac{\partial C}{\partial t} = 
    \frac{\delta}{\Stefan\Lewis}\, \frac{\partial^2 C}{\partial x^2}, 
    \label{eq:Cnondim-pde} 
  \end{gather}
  \begin{gather}
    C(x,0) = \Cinit(x) \quad \mbox{for } s_{gw}(0)<x<s_{wi}(0),
    \label{eq:Cnondim-ic}
  \end{gather}
  \begin{gather}
    C(s_{gw}(t),t) = H\rho_g(t),
    \label{eq:Cnondim-henry}
  \end{gather}
  \begin{gather}
    \frac{\partial C}{\partial x}(s_{wi}(t),t) = 0.
    \label{eq:Cnondim-bc}
  \end{gather}
\end{subequations}
Here, $\beta_w=\alpha_w\bar{t}/L^2$, $\Stefan=c_w(T_1-T_c)/\lambda$ is the Stefan
number and the Lewis number
$\Lewis=\alpha_w/D_w$ is a dimensionless ratio of thermal diffusivity of
water to the diffusivity of dissolved gas.

In the ice compartment, $s_{wi}(t)<x<1$,
\begin{subequations}\label{eq:Tinondim}
  \begin{gather}
    \frac{\partial T_i}{\partial t} =
    \beta_i \, \frac{\partial^2 T_i}{\partial x^2}, 
    \label{eq:Tinondim-pde} 
  \end{gather}
  \begin{gather}
    T_i(x,0) = T_{i0}(x) \quad \mbox{for } s_{wi}(0)<x<1,
    \label{eq:Tinondim-ic}
  \end{gather}
  \begin{gather}
    -\frac{\partial T_i}{\partial x}(1,t) = \Biot
    (T_i(1,t)-\widetilde{T}_2), 
    \label{eq:Tinondim-bc}
  \end{gather}
\end{subequations}
where $\beta_i=\alpha_i\bar{t}/L^2$ and $\widetilde{T}_2 =
(T_2-T_c)/(T_1-T_c)$.  The Biot number $\Biot={L\chtc}/{k_i}$ is a
measure of the relative resistance to heat transfer of the outer surface
of the ice to that in the interior.

The non-dimensional forms of the interfacial conditions at  
the gas-water interface are
\begin{gather}
  \frac{\partial T_g}{\partial x}(s_{gw}(t),t) = 
  \eta \frac{\partial T_w}{\partial x}(s_{gw}(t),t),
  \label{eq:match-Tg-Tw-1}
\end{gather}
\begin{gather}
  T_g(s_{gw}(t),t) = T_w(s_{gw}(t),t),
  \label{eq:match-Tg-Tw-2}
\end{gather}  
while at the water-ice interface
\begin{gather}
  T_w(s_{wi}(t),t) = T_i(s_{wi}(t),t) = 0. 
  \label{eq:match-Tw-Ti}
\end{gather}
Equation \eqref{eq:density} for the gas density reduces to
\begin{gather}
  \rho_g(t) =
  \frac{s_{gw}(0) + \int_{s_{gw}(0)}^{s_{wi}(0)} \Cinit(x)\, dx -
    \int_{s_{gw}(t)}^{s_{wi}(t)}C(x,t)\, dx}{s_{gw}(t)}.  
  \label{eq:rhognondim}
\end{gather}
The gas-water interface equation \eqref{eq:sgw-dim} becomes
\begin{gather*}
  \dot{s}_{gw}(t) = \left(1-\delta\right) \dot{s}_{wi}, 
\end{gather*}
where $\delta = \rho_i/\rho_w$, which can be integrated
directly to obtain
\begin{align}
  \left.
  \begin{array}{r}
    s_{gw}(t) = A_1 + A_2 s_{wi}(t) \\
    \text{where} \quad A_1 = s_{gw}(0) - \left(1-\delta\right) s_{wi}(0)\\
    \text{and}   \quad A_2 = 1-\delta 
  \end{array}
  \right\}.
  \label{eq:sgw-nondim}
\end{align}
Finally, the dimensionless form of the Stefan condition
\eqref{eq:stefan-dim} is
\begin{gather}
  \dot{s}_{wi} = 
  \left(\psi\frac{\partial T_i}{\partial x}(s_{wi},t) - 
    \frac{\partial T_w}{\partial x}(s_{wi},t) \right), 
  \label{eq:stefan-nondim}
\end{gather}
where $\psi=k_i/k_w$.  Note that the choice of time scale $\bar{t}$ made
in \eqref{eq:tbar} was made so that the coefficient in front of
$\dot{s}_{wi}$ scales to one.  Comparing $\bar{t}$ the typical sizes of
the corresponding scales for heat diffusion ($t_g$, $t_w$ and $t_i$
in Table~\ref{tab:2}), it is clear that the front motion occurs over a
much slower time scale.

In summary, our model consists of a coupled nonlinear system of
equations that is composed of:
\begin{itemize}
\item four PDE initial-boundary value problems
  \eqref{eq:Tgnondim}--\eqref{eq:Tinondim} for the temperatures and
  dissolved gas concentration;
\item one ODE initial value problem \eqref{eq:stefan-nondim} for the
  water-ice interface position;
\item two algebraic equations \eqref{eq:rhognondim} and
  \eqref{eq:sgw-nondim} for the gas density and gas-water interface
  position.
\end{itemize}
Because of the nonlinearity present in the equations, it is not possible
to derive an explicit analytical solution and so we must resort to
numerical simulations or approximate analytic methods.  In the next
section, we describe our numerical discretisation procedure and present
approximate results that in turn suggest an appropriate choice of
analytic solution.

\section{Numerical solution algorithm}
\label{sec:numerics}

We begin by briefly describing our approach for solving the PDEs
governing temperature and concentration.  We use the method of lines,
discretising the PDEs in space on a cell-centered grid and then solving
the resulting system of time-dependent ODEs.  In order to capture moving
boundaries sharply, we employ a moving mesh approach in which $N$
equally-spaced grid points are distributed over each of the gas, water
and ice domains, so that
\begin{alignat*}{3}
  x_g^j(t) &= (j-1/2) h_g(t) 
  &&\qquad \text{with } h_g(t) = \frac{s_{gw}(t)}{N}, \\
  x_w^j(t) &= s_{gw}(t) + (j-1/2) h_w(t) 
  &&\qquad \text{with } h_w(t) = \frac{s_{wi}(t)-s_{gw}(t)}{N}, \\
  x_i^j(t) &= s_{wi}(t) + (j-1/2) h_i(t) 
  &&\qquad \text{with } h_i(t) = \frac{1-s_{wi}(t)}{N},
\end{alignat*}
for $j=1, 2, \dots, N$, and where $h_\ell(t)$ for $\ell=g,w,i$ denotes the grid spacing on the
corresponding compartment.  When using such a moving computational grid,
we must introduce an additional convection term in each parabolic PDE
owing to the grid motion~\cite{MR605400,MR2722625}  
\begin{gather}
  \frac{\partial f}{\partial t} - \meshvel\, 
  \frac{\partial f}{\partial x} 
  = \kappa\,\frac{\partial^2 f}{\partial x^2} , 
  \label{eq:moving}
\end{gather}
where $f=T_g$, $T_w$, $T_i$, $C$ and $\kappa=\beta_g$, $\beta_w$,
$\beta_i$, $\delta/(\Stefan\Lewis)$ respectively.  The convective term
has a velocity $\meshvel$ that corresponds to the mesh velocity
$\dot{x}_\ell$ on the gas and ice compartments, but equals
$\dot{x}_w-\dot{s}_{gw}$ on the water compartment owing to the presence
of the convective term in \eqref{eq:Twnondim-pde}.

The spatial derivatives appearing in equation \eqref{eq:moving} are
replaced using centered, second-order difference approximations to
obtain
\begin{gather}\label{eq:moving_discr}
  \frac{\partial f^j_\ell}{\partial t} -
  \meshvel^j_\ell\, \frac{f^{j+1}_\ell-f^{j-1}_\ell}{2h_\ell} = 
  \kappa_\ell\, \frac{f^{j+1}_\ell-2f^{j}_\ell+f^{j-1}_\ell}{h_\ell^2},
\end{gather}
where $f^j_\ell(t) \approx f(x_\ell^j,t)$ are the discrete
approximations of the dependent variables for $\ell=g,w,i$ and
$j=1,2,\dots,N$.  Centered finite differences are also used to
discretise the boundary conditions and hence maintain second order
accuracy throughout.  This requires values of the approximate solution
that lie within a grid cell lying immediately outside of each
compartment; to this end we introduce fictitious points $x_\ell^0 =
x_\ell^1 - h_\ell$ and $x_\ell^{N+1} = x_\ell^N + h_\ell$.  A Dirichlet
boundary condition such as \eqref{eq:Tgnondim-bc} is
approximated using an arithmetic average
\begin{gather*}
  \frac{T_g^0 + T_g^1}{2} = 1,  
\end{gather*}
which is solved for the fictitious value as $T_g^0=2-T_g^1$.
Furthermore, a Neumann boundary condition such as \eqref{eq:Tinondim-bc}
is approximated by
\begin{gather*}
  - \frac{T_i^{N+1} - T_i^{N}}{h_i} = \Biot \left( \frac{T_i^N +
      T_i^{N+1}}{2} - \widetilde{T}_2 \right), 
\end{gather*}
which yields
\begin{gather*}
  T_i^{N+1} = \frac{(2-h_i\Biot)T_i^N + 2h_i\Biot
    \widetilde{T}_2}{2+h_i\Biot} . 
\end{gather*}
The remaining fictitious point values are obtained in a similar manner
using the other boundary and matching conditions.  Finally, integrals of
concentration that appear in the boundary condition
\eqref{eq:Cnondim-henry} (via the density \eqref{eq:rhognondim}) are
approximated using the trapezoidal rule, so that the resulting spatial
discretisation is fully second order in space.

The semi-discrete temperature and concentration equations comprise a
system of $4N$ time-dependent ODEs.  One additional ODE derives from the
water-ice interface equation \eqref{eq:stefan-nondim} in which the
spatial derivatives are also approximated using centered differences.
The resulting system of $4N+1$ ODEs is implemented in the
Matlab$^{\mbox{\textregistered}}$ programming environment and integrated
in time using the stiff solver \texttt{ode15s}.  In all cases, the error
tolerances for \texttt{ode15s} are set to \texttt{AbsTol=1e-10} and
\texttt{RelTol=1e-8}.

\section{Numerical simulations}
\label{sec:results}

The method described in the previous section is now employed to simulate
the model equations and to evaluate its sensitivity to various
physical parameters.  In all simulations, we make the following choices
for initial conditions:
\begin{itemize}
\item $s_{gw}(0)=0.1$ and $s_{wi}(0)=0.11$, so that the water is
  initially completely frozen except for a thin liquid layer (refer to
  Assumption \ref{assume:waterlayer});
\item $C(x,0)\equiv 0$, corresponding to no dissolved gas;
\item $T_g(x,0)\equiv T_w(x,0)\equiv 1$ and $T_i(x,0)\equiv
  \widetilde{T}_2$, so that the gas and water compartments are both
  equilibrated with the left boundary temperature, while the ice
  compartment is equilibrated with the ambient (sub-freezing)
  temperature at the right boundary.
\end{itemize}

\begin{figure}
  \footnotesize
  \centering
  \begin{tabular}{cc}
    (a) Gas concentration at $x=L$. & 
    (b) Gas concentration profiles (long time). \\
    \includegraphics[width=0.45\textwidth]{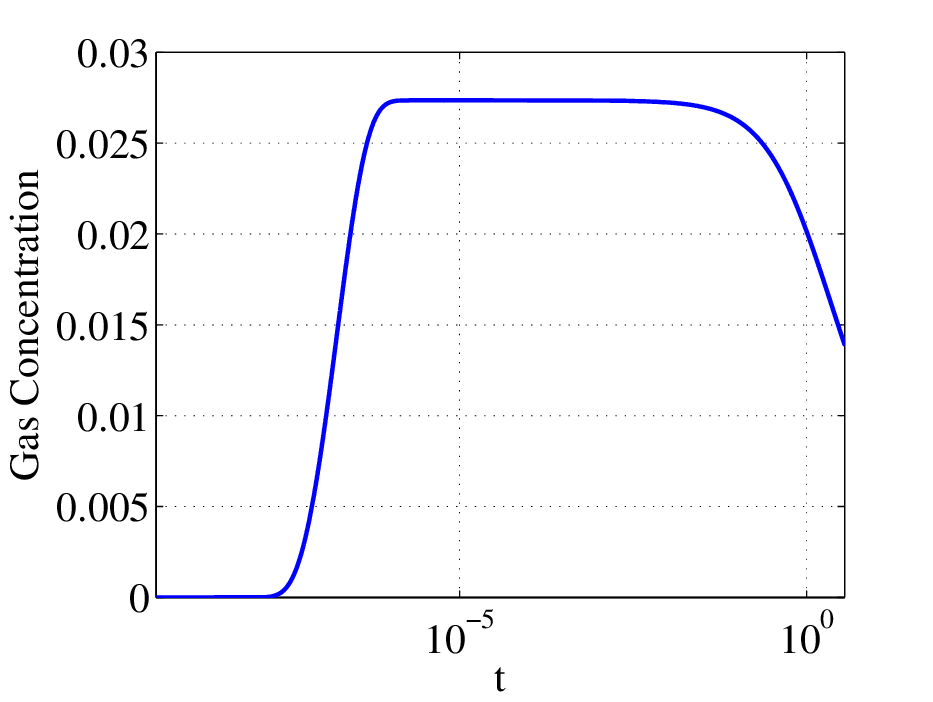} & 
    \includegraphics[width=0.45\textwidth]{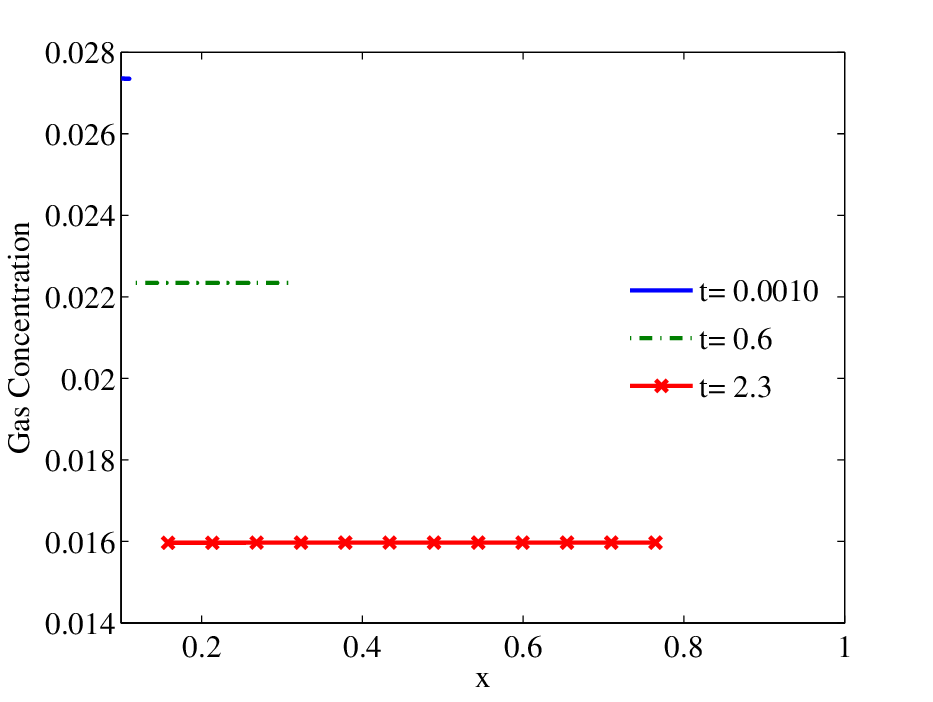}\\
    (c) Concentration profiles (short time). &
    (d) Phase interfaces. \\
    \includegraphics[width=0.45\textwidth]{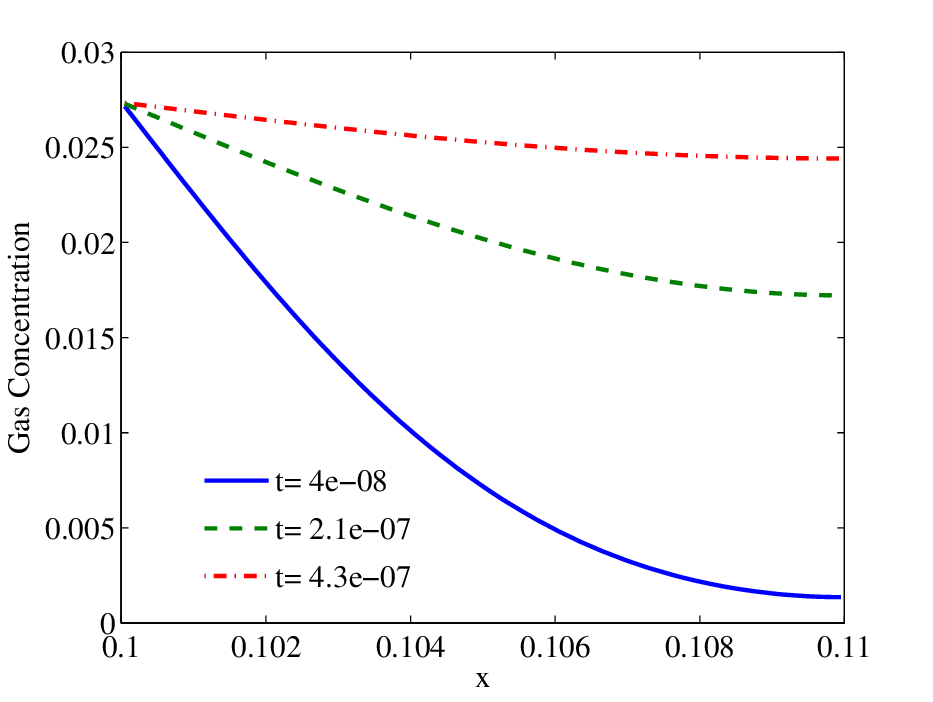} & 
    \includegraphics[width=0.45\textwidth]{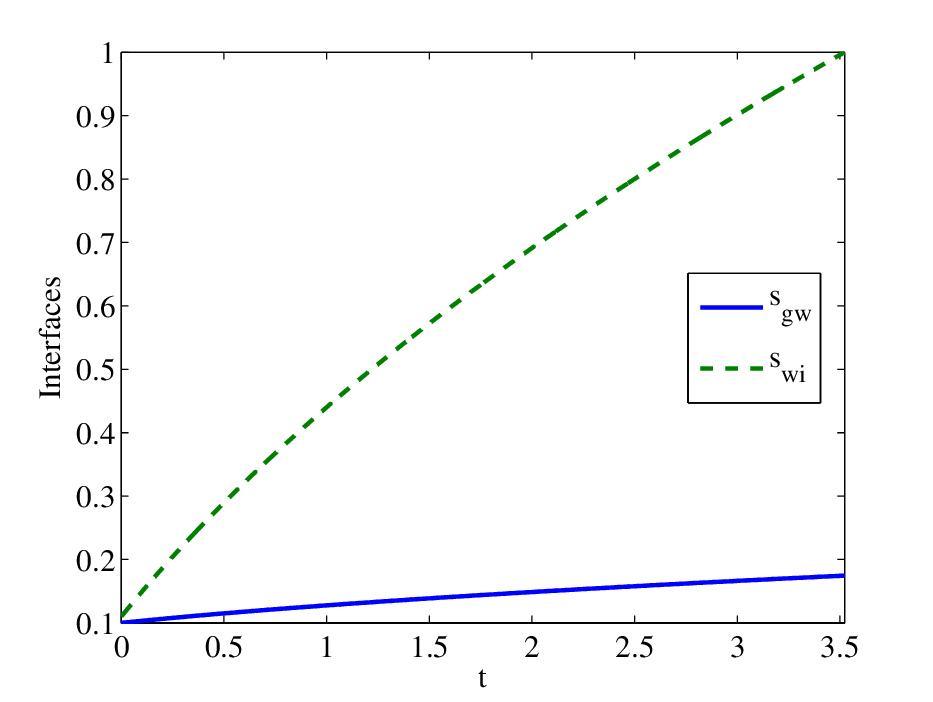}\\
    (e) Temperatures. & \\
    \includegraphics[width=0.45\textwidth]{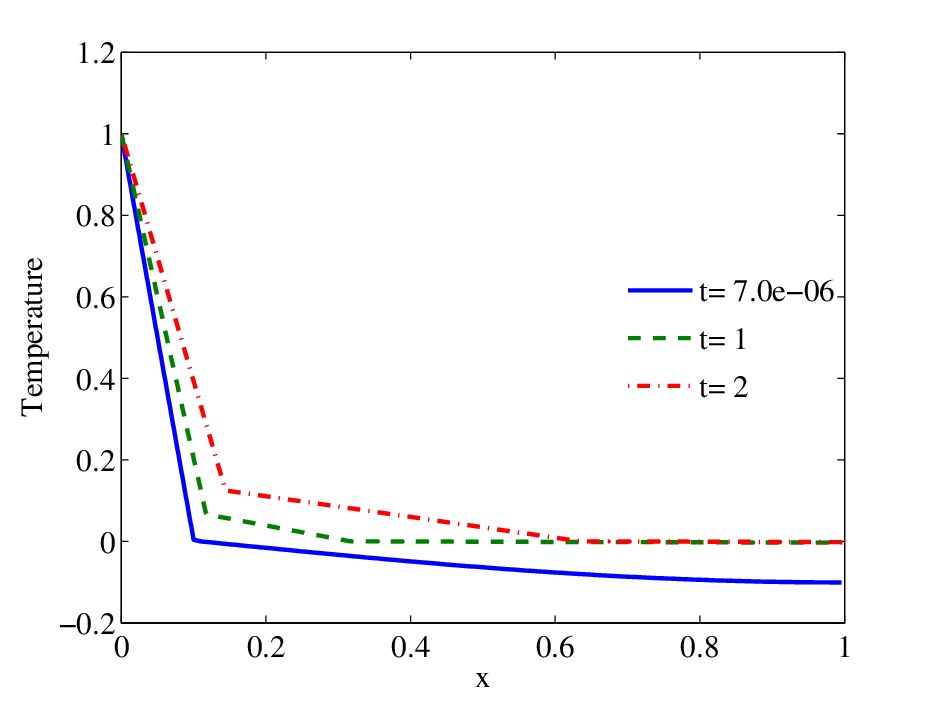} & 
  \end{tabular}
  \caption{Solution plots with boundary temperatures $T_1=T_c+0.005$ and
    $T_2=T_c-0.005$, with $\tilde{T}_2=-1$.}
  \label{fig:2}
\end{figure}

\begin{figure}
  \footnotesize
  \centering
  \begin{tabular}{cc}
    (a) Gas concentration at $x=L$. & 
    (b) Gas concentration profiles (long time). \\
    \includegraphics[width=0.45\textwidth]{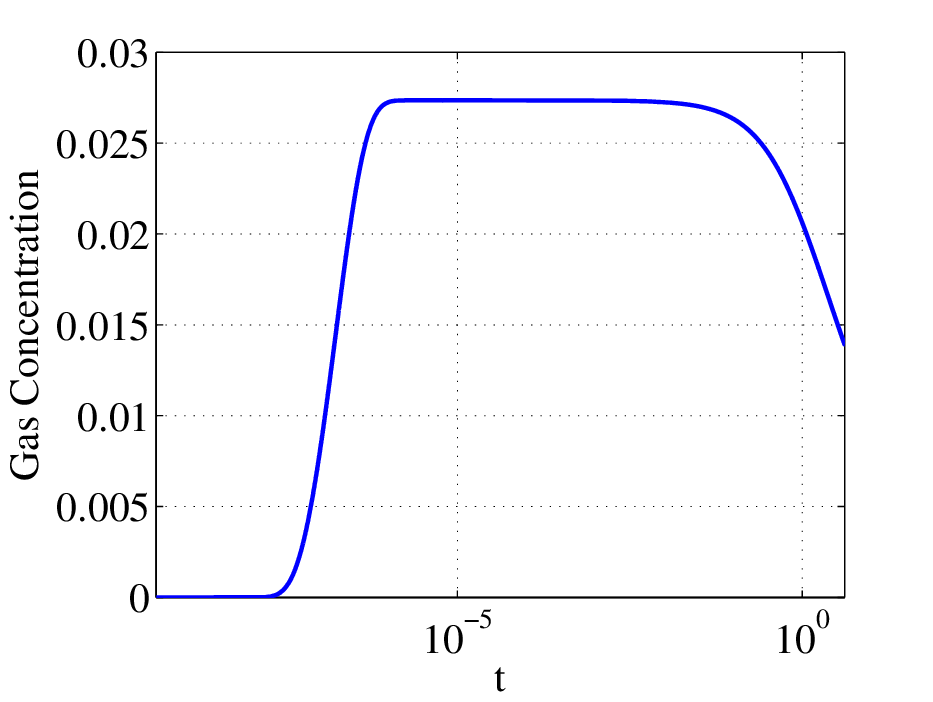} & 
    \includegraphics[width=0.45\textwidth]{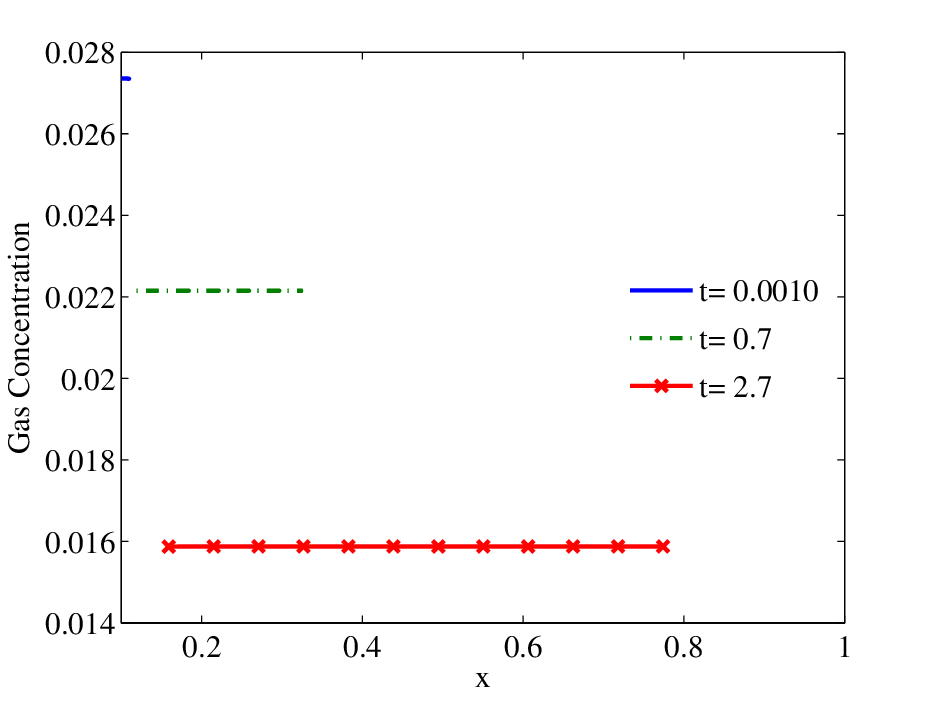}
  \end{tabular}
  \caption{Solution plots with boundary temperatures $T_1=T_c+0.005$ and
    $T_2=T_c-0.02$, with $\tilde{T}_2=-4$.} 
  \label{fig:3}
\end{figure}

We begin by focusing on the water-ice front motion that drives the phase
change dynamics and in turn influences the gas dissolution.  We
consider as a ``base case'' the situation where the left and right
boundary temperatures are $T_1=T_c+0.005$ and $T_2=T_c-0.005$, for which
results are provided in Figure~\ref{fig:2}.  Two other cases with larger
values of $T_1$ and $T_2$ are presented in
Figures~\ref{fig:3} and \ref{fig:41} for comparison purposes.   Note
that all plots are show in dimensionless variables.

\leavethisout{
\begin{figure}
  \footnotesize
  \centering
  \begin{tabular}{cc}
    (a) Gas concentration at $x=L$. & 
    (b) Gas concentration profiles (long time). \\
    \includegraphics[width=0.45\textwidth]{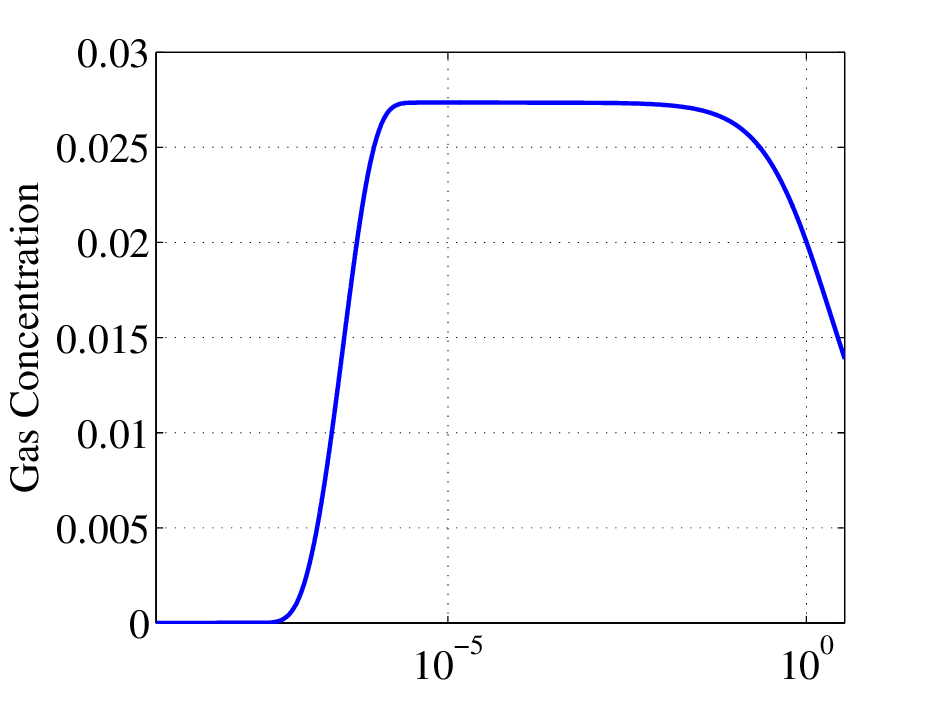} & 
    \includegraphics[width=0.45\textwidth]{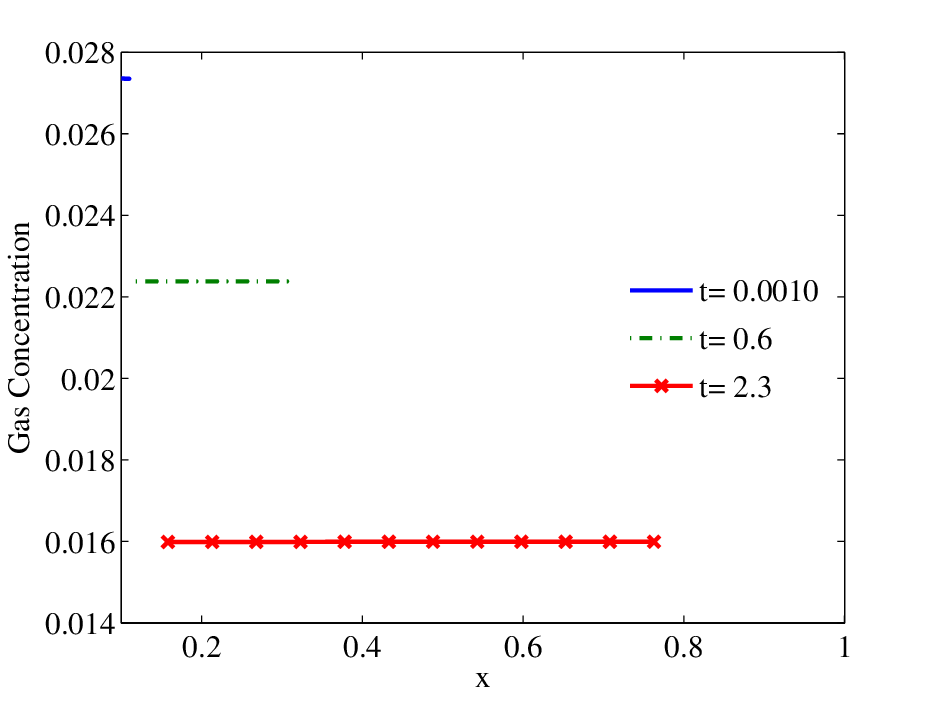}
  \end{tabular}
  \caption{Solution plots with boundary temperatures $T_1=T_c+0.01$ and
    $T_2=T_c-0.005$, $\tilde{T}_2=-0.5$.}
  \label{fig:4}
\end{figure} 
}

\begin{figure}
  \footnotesize
  \centering
  \begin{tabular}{cc}
    (a) Gas concentration at $x=L$. & 
    (b) Gas concentration profiles (long time). \\
    \includegraphics[width=0.45\textwidth]{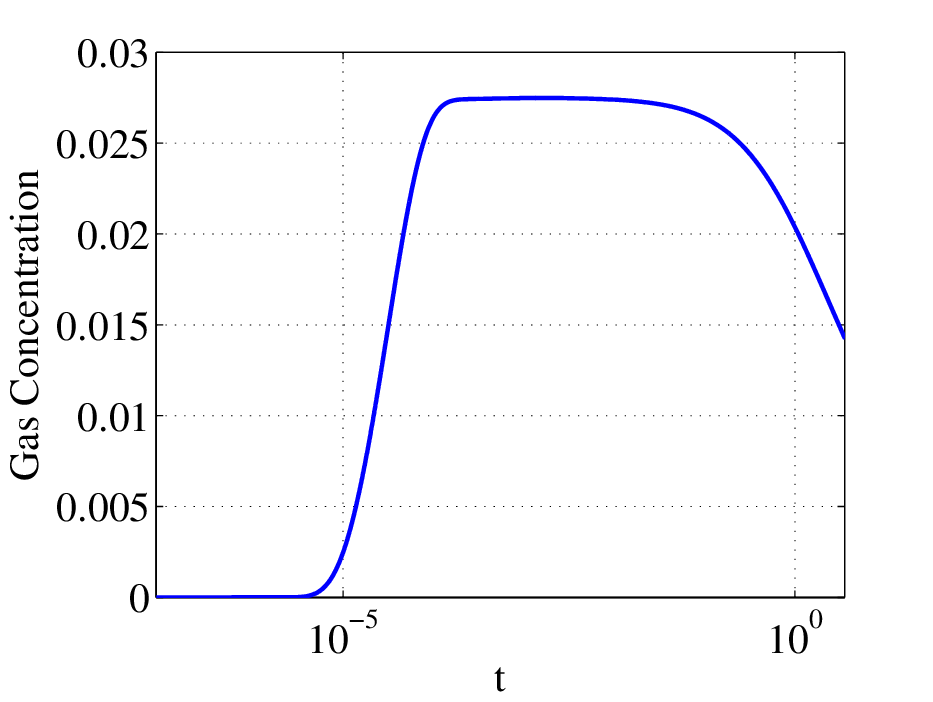} & 
    \includegraphics[width=0.45\textwidth]{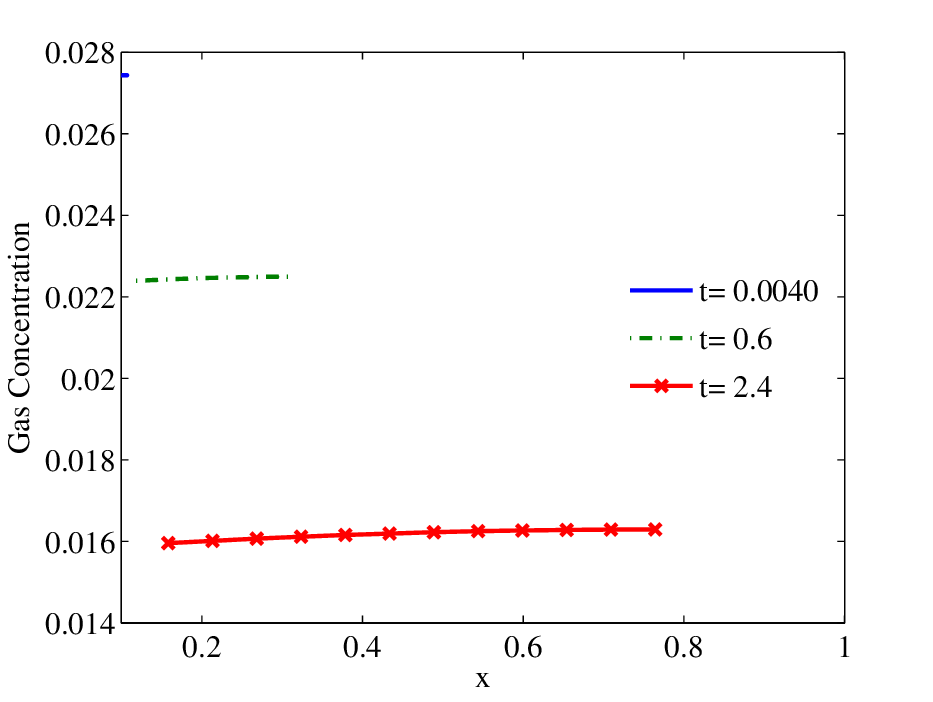}\\
  \end{tabular}
  \caption{Solution plots with boundary temperatures $T_1=T_c+1$ and
    $T_2=T_c-1$, $\tilde{T}_2=-1$.}
  \label{fig:41}
\end{figure} 

The the base case, the plot in Figure~\ref{fig:2}a of the dissolved gas
concentration (measured at the right-hand boundary) exhibits a clear
division of the solution behaviour into three separate time periods:
\begin{enumerate}
\item A very short initial transient during which the concentration
  undergoes a rapid increase from zero at $t=0$ to some maximum value
  at $t\sim O(10^{-7})$.  This transition layer arises because we have
  chosen initial conditions corresponding to zero dissolved gas and
  hence Henry's law forces the initially very thin water layer to
  rapidly ``fill up'' with gas.  The corresponding diffusion
  of dissolved gas within the water compartment is easily seen in
  Figure~\ref{fig:2}c. 

\item The gas concentration remains roughly unchanged over the interval
  $t\in[10^{-7},10^{-2}]$, since the water does not yet melt
  appreciably.

\item The time $t \approx 10^{-2}$ signals the onset of ice melting,
  after which the water compartment begins to grow in size.  Even though
  this allows more gas to dissolve in the water layer, the increased
  volume leads to a decrease in the dissolved gas concentration as the
  pressure in the gas compartment decreases.  This effect is evident
  from Figure~\ref{fig:2}b, where we observe that the concentration
  profiles through the water layer are roughly constant in $x$, although
  there is a very slight increase in $C$ from left to right.
\end{enumerate}
The presence of these three, clearly separated time scales is a
characteristic feature of the evolution of dissolved gas.  Because the
concentration profiles are almost constant in $x$ over longer times, the
gas concentration dynamics are driven primarily by the relative motion
of the free boundaries.

A somewhat counter-intuitive result derives from the observation that
after initial transients are complete, $C$ attains its maximum value at
the water-ice interface rather than at the gas-water interface where
dissolution is actually taking place.  This slight positive slope in the
plot of $C$ versus $x$ becomes more pronounced as the boundary
temperature difference $T_1-T_2$ is increased, and can be seen most
clearly in Figure~\ref{fig:41}b where the temperature difference is
largest.  We also remark that the speed of the free boundaries increases
with $T_1-T_2$ which allows less time for the gas to adjust in the water
compartment.

We close our discussion of the base case with a look at the final two
plots in Figure~\ref{fig:2}.  The water-ice interface in
Figure~\ref{fig:2}d shows the expected sub-linear behaviour that is
consistent with the ${t}^{1/2}$ dependence predicted by the analytical
solution to the Stefan problem.  This behaviour is confirmed by our
asymptotic results in Section~\ref{sec:asy-swi}.  According to
Figure~\ref{fig:2}e, the temperature field is a continuous function that
changes relatively slowly over time.  Furthermore, the temperature is
approximately linear within each compartment, with a pronounced ``kink''
at the each interface locations.  Both of these results will be explained
by the analytical solution derived in Section~\ref{sec:asymptotics}.

The effect of increasing the temperature difference $T_1-T_2$ can be
seen by comparing the results in Figures~\ref{fig:3} and \ref{fig:41}
with Figure~\ref{fig:2}.  There is a significant slowing of the initial
transient gas dissolution dynamics as $T_1-T_2$ is increased, although
the long-time concentration dynamics are largely unchanged.  However, as
mentioned above, there is a slight increase in the slope of the
concentration profiles in Figure~\ref{fig:41}b.

Comparing Figures~\ref{fig:2}--\ref{fig:41}, we remark that for the base
case with temperature increments of $0.005$, the ice layer melts away
after about one hour.  In contrast, the melting time shortens to 17
seconds when the temperature increment is taken as large as 1.0.  The
only place that $\bar{T}_1$ enters the model is through the Stefan number
$\Stefan$, which explains why changes in $\bar{T}_1$ have the effect of
altering the time scale for the free boundary motion.

\begin{figure}
  \footnotesize
  \centering
  \begin{tabular}{cc}
    (a) Concentration profiles (short time). & (b) Gas concentration at $x=L$. \\
    \includegraphics[width=0.45\textwidth]{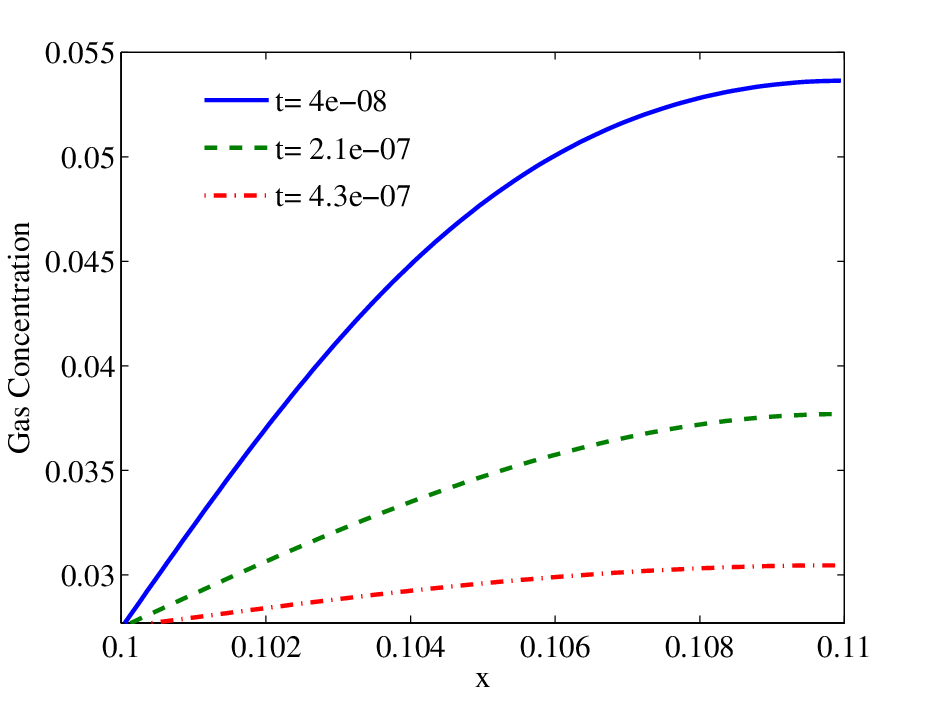}
    &
    \includegraphics[width=0.45\textwidth]{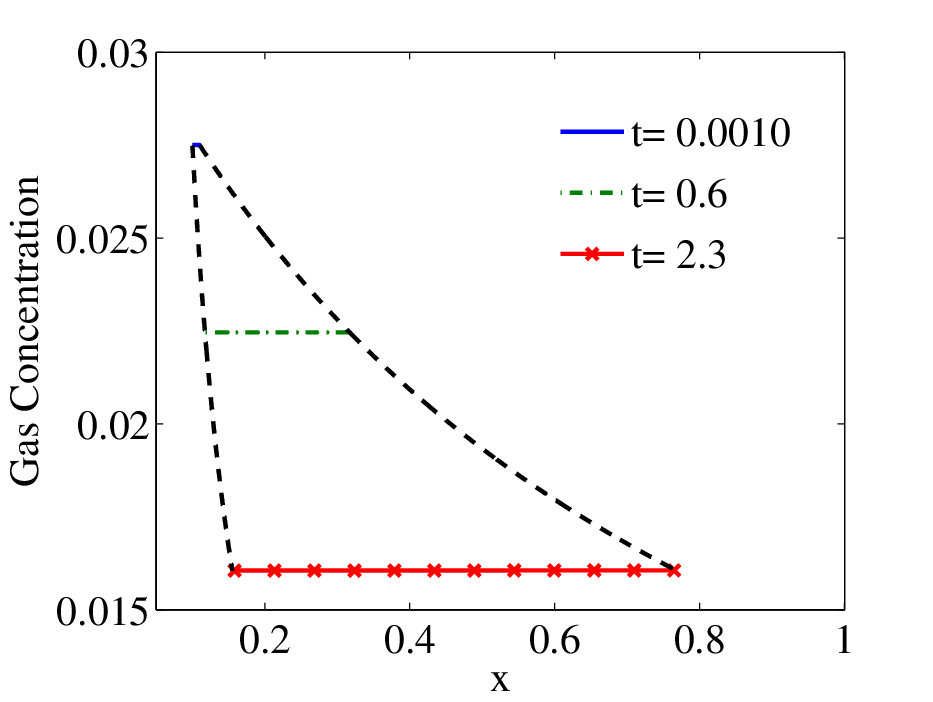}
  \end{tabular}
  \caption{Comparison of analytical and numerical solutions with
    boundary temperatures $T_1=T_c+0.005$, $T_2=T_c-0.005$,
    $\tilde{T}_2=-1$, and $\Cinit(y)=2\bar{C}$.}  
  \label{fig:10}
\end{figure}

We conclude this section by investigating the effect of taking a
relatively large initial value for the dissolved gas concentration,
$C(x,0)\equiv 0.055$, rather than taking $C(x,0)\equiv 0$ as we have so
far.  We will see later on that this initial concentration is large in
the sense that it is twice the steady-state value of concentration for
the base case.  Hence, this situation may be viewed as corresponding to
a ``super-saturated'' case in which one would expect dissolved gas to
immediately cavitate and form bubbles. The results in
Figure~\ref{fig:10}a are consistent with this hypothesis, and show that
the behaviour of the concentration profiles is reversed relative to the
base case in Figure~\ref{fig:2}c, in that concentration decreases from
the initial value to its quasi-steady state.

\section{Approximate analytical solutions}
\label{sec:asymptotics}

Motivated by the numerical results in the previous section, we now
derive an approximate analytical solution that is based on the following
observations:
\begin{itemize}
\item The temperature is approximately linear within each compartment,
  and equilibrates rapidly to any change in conditions over the time
  scale of the interface motion.  This suggests using a quasi-steady
  approximation for each temperature variable.
\item The dissolved gas concentration evolves over two distinct time
  scales: a rapid initial equilibration phase driven by diffusion (on
  the order of $10^{-8}$--$10^{-5}$ seconds) during which gas dissolves
  at the gas-water interface to fill the liquid compartment; and a much
  longer time scale corresponding to the onset of ice melting (on the
  order of $t=0.01$--0.1\;\units{s}) when the water-ice interface begins
  to move and the volume of the water compartment increases appreciably.
\end{itemize}
As a result, we approximate the solution in three stages.  First, we
make a quasi-steady approximation for temperature that permits us to
write $T_\ell(x,t)$ as linear functions of $x$ for $\ell=g,w,i$, that
vary in time only through changes in the interface locations.  Second,
we derive a simpler ODE for the water-ice interface $s_{wi}(t)$ that
makes use of a series expansion in the small parameter $\Biot$, which
then also yields an approximation for $s_{gw}(t)$ via equation
\eqref{eq:sgw-nondim}.  Finally, we develop a two-layer asymptotic
solution for the dissolved gas concentration $C(x,t)$ based on the
separation of time scales mentioned above.

\subsection{Quasi-steady approximation for temperatures}
\label{sec:temperatures}

The time scales for diffusion of heat in the gas, water and ice
compartments can be estimated using
\begin{gather*}
  t_\ell = \frac{L^2}{\alpha_\ell} \qquad \mbox{for } \ell=g, w, i, 
\end{gather*}
where the thermal diffusivities $\alpha_\ell$ and length scale $L$ are
taken from Tables~\ref{tab:1} and~\ref{tab:2}.  The corresponding time
scales are $t_g\approx 5.4 \times 10^{-4}\;\units{s}$, $t_w\approx 7.2
\times 10^{-2}\;\units{s}$, and $t_i\approx 8.5 \times
10^{-3}\;\units{s}$.  In contrast, the time scales for motion of the
gas-water and water-ice interfaces were observed in the numerical
simulations from the previous section to be at least one order of
magnitude larger than this; consequently, the phase temperatures will
adjust rapidly in response to any motion of the interfaces.  It is
therefore reasonable to assume that the temperatures $T_\ell$ are
quasi-steady in the sense that they do not depend explicitly on time but
instead have an implicit dependence on $t$ through the free boundary
locations $s_{gw}(t)$ and $s_{wi}(t)$.

The convective term in the water equation \eqref{eq:Twnondim-pde} is so
small (on the order of $10^{-2}$) that it is reasonable
to neglect.  Therefore the temperature equation in all three
compartments has the simple form $\partial_{xx} T_\ell=0$ and
consequently the temperature is well-approximated by linear functions of
$x$
\begin{gather}
  T_\ell(x,t) = a_\ell(t) x + b_\ell(t) \qquad \mbox{for } \ell=g,w,i. 
  \label{simplyT1}
\end{gather}
The coefficients $a_\ell(t)$ and $b_\ell(t)$ can be determined by
imposing boundary and matching conditions \eqref{eq:Tgnondim-bc},
\eqref{eq:Tinondim-bc} and
\eqref{eq:match-Tg-Tw-1}--\eqref{eq:match-Tw-Ti}, after which we obtain
\begin{subequations}\label{eq:Texact}
  \begin{gather}
    T_g(x,t) = - \frac{\eta (x - s_{gw}) + s_{gw} - s_{wi}}{s_{wi} +
      (\eta-1) s_{gw}},
    \label{eq:Tg-exact}
  \end{gather}
  \begin{gather}
    T_w(x,t) = \frac{s_{wi}-x}{s_{wi}+(\eta-1)s_{gw}}, 
    \label{eq:Tw-exact}
  \end{gather}
  \begin{gather}
    T_i(x,t) = \frac{\Biot \widetilde{T}_2(x-s_{wi})}{1+\Biot(1-s_{wi})},
    \label{eq:Ti-exact}
  \end{gather} 
\end{subequations}
on the corresponding sub-intervals.

\subsection{Asymptotic expansion for water-ice interface}
\label{sec:asy-swi}

We next derive an analytical solution for the water-ice interface
$s_{wi}(t)$ by substituting the approximations just derived for $T_i$
and $T_w$ into the Stefan condition \eqref{eq:stefan-nondim} along with
the expression \eqref{eq:sgw-nondim} for $s_{gw}$ to obtain the
following ODE
\begin{gather}
  (B_1+B_2s_{wi})(1+\Biot(1-s_{wi}))\dot{s}_{wi} =
  (1+\Biot(B_5+B_6s_{wi})). 
  \label{stefan_an}
\end{gather}
The constants appearing in this equation are
\begin{gather*}
  \begin{array}{ll}
    B_1 = (\eta-1)A_1,
    &    B_2 = 1+(\eta-1)A_2,\\
    B_3 = B_1 s_{wi}(0) + B_2 \frac{s_{wi}^2(0)}{2},
    &    B_4 = B_1 s_{wi}(0) + \frac{B_2-B_1}{2} s_{wi}^2(0) -
               \frac{B_2}{3}s_{wi}^3(0),\\ 
    B_5 = 1 + \psi\widetilde{T}_2 B_1,
    &    B_6 = \psi\widetilde{T}_2 B_2 - 1,
  \end{array}
\end{gather*}
while $A_1$ and $A_2$ are the same constants defined earlier in equation
\eqref{eq:sgw-nondim}.  This ODE can be integrated in time over the
interval $[0,t]$ to obtain the following integral equation for $s_{wi}$:
\begin{multline}
  \label{stefan_int}
  B_1 s_{wi} + \frac{B_2}{2} s_{wi}^2 + \Biot \left[ B_1 s_{wi} +
    \frac{B_2-B_1}{2} s_{wi}^2 - \frac{B_2}{3} s_{wi}^3 \right]\\
  = B_3 +  t + \Biot B_4 + \Biot
   \left[ B_5 t + B_6 \int_0^t s_{wi}(l)\, dl
  \right]. 
\end{multline}
Because the Biot number satisfies $\Biot\ll 1$ (see Table~\ref{tab:2})
it is reasonable to look for a series solution of the form
\begin{subequations}\label{eq:swi-asy}
  \begin{gather}
    s_{wi}(t) = s_0(t) + \Biot s_1(t) + O(\Biot^2).
    \label{eq:swi-series}
  \end{gather}
  Substituting this expression into \eqref{stefan_int} and collecting
  terms in like powers of $\Biot$, we find that to leading order
  \begin{align}
    s_0(t) &= \frac{1}{B_2} \left( \sqrt{B_1^2 + 2B_2 \left( B_3 +
           t \right)} - B_1 \right),
    \label{eq:swi-s0}\\
    \intertext{while the next order correction is}
    s_1(t) &=
    \frac{1}{B_1 + B_2 s_0(t)} \left( \frac{B_2}{3} s_0(t)^3 +
      \frac{B_1-B_2}{2} s_0(t)^2 - B_1 s_0(t) + B_4 \right.
    \nonumber \\
    & \hspace*{3cm} 
    + \,  B_5 t + \left. B_6
      \int_0^t s_0(l)\, dl \right).
    \label{eq:swi-s1}
  \end{align}
\end{subequations}
Using the water-ice interface approximation in equations
\eqref{eq:swi-asy} the gas-water interface may be determined from
\eqref{eq:sgw-nondim}.

\begin{figure}
  \centering
  \includegraphics[width=0.65\textwidth]{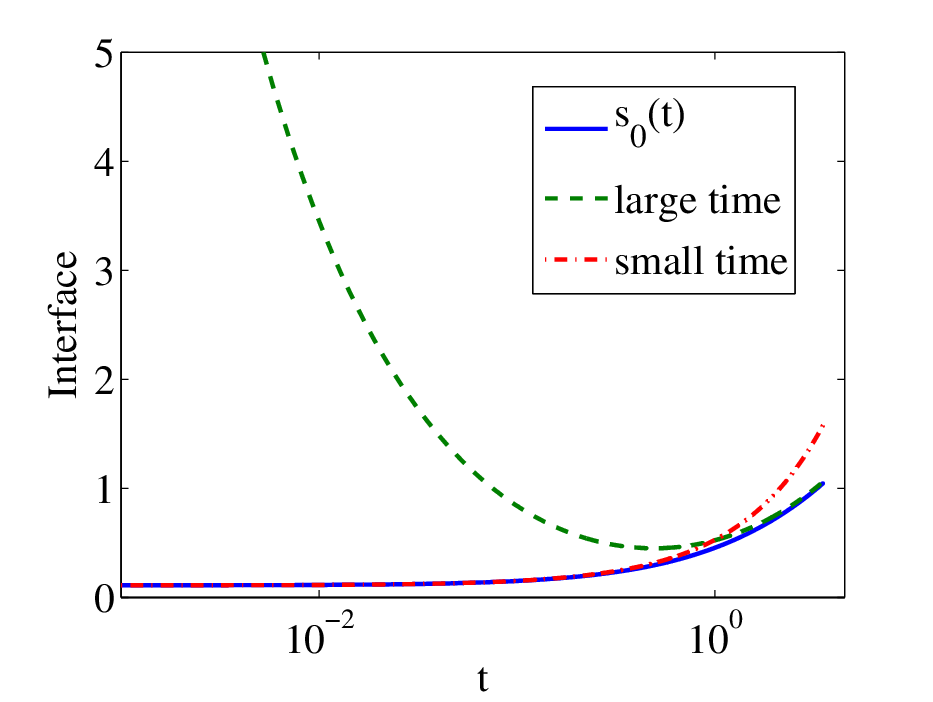}
  \caption{Series expansions of $s_0(t)$ for large and small times,
    showing the expected $\sqrt{t}$ behaviour as $t\rightarrow \infty$
    (note that the $t$-axis is on a log scale).}
  \label{fig:s0-series}
\end{figure}

We conclude this section by drawing a connection between the leading
order solution $s_0(t)$ in the limit as $\Biot\rightarrow 0$ and the
classical solution of the Stefan problem where the melting front moves
with a speed proportional to $t^{1/2}$.  Although equation
\eqref{eq:swi-s0} does not have exactly this form, the behaviour is
consistent in the limits of large and small time.  In particular, if we
expand \eqref{eq:swi-s0} in a Taylor series about $t=0$ we find that
\begin{gather}
  s_0(t) = \frac{\sqrt{B_1^2 + 2B_2 B_3} - B_1}{B_2} +
  \frac{2}{\sqrt{B_1^2 + 2B_2 B_3}} \, t + O(t^2) 
  \qquad \text{(as $t\rightarrow 0$)}.
  \label{eq:s0-taylor}
\end{gather}
Furthermore, the large-time limit of \eqref{eq:swi-s0} yields 
\begin{gather}
  s_0(t) = \frac{-B_1}{B_2} + \left( \frac{2t}{B_2} \right)^{1/2} + 
  \frac{B_1^2 + 2B_2 B_3}{2B_2^2} \, \left( \frac{2t}{B_2}
  \right)^{-1/2} + O(t^{-3/2}) \qquad \text{(as $t\rightarrow \infty$)}.
  \label{eq:s0-asy}
\end{gather}
When these two series expansions are plotted against the exact expression
for $s_0(t)$ in Figure~\ref{fig:s0-series}, we see that both match 
well for small and large times, and in particular the large-time
expansion \eqref{eq:s0-asy} shows the expected $t^{1/2}$ behaviour.

\subsection{Two-scale asymptotic solution for gas concentration} 
\label{sec:asy-c}

The numerical simulations from Section \ref{sec:results} (more
specifically, the plots in Figures~\ref{fig:2}a, \ref{fig:3}a,
\ref{fig:41}a) exhibited a clear separation of time scales during the
evolution of the dissolved gas concentration.  Starting from the given
initial value, the concentration increases rapidly as gas dissolves at
the gas-water interface and diffuses throughout the water compartment.
We repeat our earlier observation that the gas concentration is nearly
constant in space, but has a slight positive slope that leaves the
maximum value at the water-ice interface (see Figure~\ref{fig:41}b);
this maximum is achieved over the short diffusion time scale before the
free boundaries begin to move.  From then on, the dissolved gas
concentration remains essentially linear and decreases over a much
longer time scale that is driven by the motion of the free boundaries.
It is this dual time scale behaviour that we aim to explain in this
section.

\leavethisout{%
  In fact, see \cite[sect. 3.32]{Crank1956}, the time needed by the
  diffusion front to travel a distance $d$ is given by:
  \begin{gather*}
    t_d=\frac{d^2}{4D_w}.
  \end{gather*}
}

To this end, it is helpful to derive rough estimates of the time and
length scales involved.  The time required for the dissolved gas to
diffuse a distance $d=\frac{L}{2}(s_{wi}(0)-s_{gw}(0))$ corresponding to
half the width of the water compartment is
\begin{gather}
  t_d = \frac{d^2}{D_w}
  \approx 1.13\times 10^{-2}\;\units{s}.
  \label{eq:tdiff}
\end{gather}
This value should be compared with the time $t_{wi}$ required for the
ice to melt completely, which can be estimated by setting
$s_{wi}(t_{wi}) = 1$ in equation \eqref{eq:swi-asy} and focusing on the
leading order term to obtain
\begin{gather*}
  t_{wi} = \frac{\bar{t}}{2}(B_2+2B_1-2B_3) 
  \approx 96.4~\units{hours},
\end{gather*}
which is six orders of magnitude larger than the diffusion scale $t_d$
in \eqref{eq:tdiff} above.  Moreover, over this same time scale, the
water-ice interface is only capable of travelling a distance of
\begin{gather*}
  L(s_{wi}(t_d)-s_{wi}(0)) \approx 2.25\times 10^{-8} L.
\end{gather*}
Hence, the phase interfaces can certainly be treated as stationary over
the diffusive time scale $t_d$.

Based on these observations, we now develop a two-layer asymptotic
expansion for the dissolved gas concentration.\ \ 
We begin by rescaling the dimensionless spatial variable according to
\begin{gather}
  y = \frac{x-s_{gw}(0)}{\Delta s}, 
  \label{eq:y}
\end{gather}
where $\Delta s = s_{wi}(0)-s_{gw}(0)>0$.  By substituting into equation
\eqref{eq:Cnondim-pde} and defining a new concentration variable
$G(y,t) = C(x,t)$, we obtain
\begin{gather}
  \frac{\partial G}{\partial t} =
  \frac{1}{\epsilon} \frac{\partial^2 G}{\partial y^2}, 
  \label{adimc:1-2}
\end{gather}
where the new diffusion parameter is
\begin{gather*}
  \epsilon = \frac{\Lewis\Stefan}{\delta} (\Delta s)^2 \ll 1.
\end{gather*}
It is convenient at this point to rescale the interface positions
according to
\begin{gather*}
  \sigma_{wi}(t) = \frac{s_{wi}(t)-s_{gw}(0)}{\Delta s} \qquad
  \text{and} \qquad
  \sigma_{gw}(t) = \frac{s_{gw}(t)-s_{gw}(0)}{\Delta s}.
\end{gather*}
Two series expansions will next be developed for the concentration
variable $G(y,t)$: one on an ``outer region'' corresponding to times
$t=O(1)$, and the second on an ``inner region'' corresponding to
$t=O(\epsilon)\ll 1$.

\subsubsection{Outer expansion (large time)}

For large times, we suppose that the dissolved gas concentration is a
series in the small parameter $\epsilon$:
\begin{gather}
  \Couter(y,t) = \Couter_0(y,t) + \epsilon \Couter_1(y,t) +
  O(\epsilon^2). 
  \label{eq:outer}
\end{gather}
Substituting this expression into \eqref{adimc:1-2} and collecting terms
with like powers of $\epsilon$ gives rise to the leading order equation
\begin{align*}
  \frac{\partial^2 \Couter_0}{\partial y^2} = 0,
\end{align*}
which has solution $G(y,t) = a(t) y + b(t)$, similar to the quasi-steady
approximation for temperature we obtained in
Section~\ref{sec:temperatures}.  The leading order boundary conditions
corresponding to \eqref{eq:Cnondim-henry} and \eqref{eq:Cnondim-bc} are
\begin{align*}
  \displaystyle
  \frac{\partial \Couter_0}{\partial y} (\sigma_{wi}(t),t) &= 0,\\
  \Couter_0(\sigma_{gw}(t),t) &= \frac{H \zeta + H \left(
      \ds{\int_{0}^{1} \Cinit(y)\, dy} -
      \ds{\int_{\sigma_{gw}(t)}^{\sigma_{wi}(t)} \Couter_0(y,t)\, dy} \right)}{\zeta
    + \sigma_{gw}(t)}, 
\end{align*}
where we have introduced the notation
\begin{gather}
  \zeta = \frac{s_{gw}(0)}{\Delta s}, 
  \label{eq:zeta}
\end{gather}
which is a positive constant because $\Delta s>0$ by
Assumption~\ref{assume:waterlayer}.  The zero Neumann boundary condition
requires that $a(t)\equiv 0$, after which we obtain the leading order
solution
\begin{gather}
  \Couter_0(y,t) =  \frac{H \zeta + H \ds{\int_{0}^{1}
      \Cinit(y)\, dy}}{\zeta + \sigma_{gw} + H (\sigma_{wi}-\sigma_{gw})}. 
  \label{eq:outer0}
\end{gather}

At the next higher order in $\epsilon$, we obtain the following boundary
value problem for $G_1(y,t)$
\begin{align*}
  \frac{\partial^2 \Couter_1}{\partial y^2} &= \frac{\partial
    \Couter_0}{\partial t},\\
  \frac{\partial \Couter_1}{\partial y}(\sigma_{wi}(t),t) &= 0,\\
  \Couter_1(\sigma_{gw}(t),t) &= -H \xi(t) \int_{\sigma_{gw}(t)}^{\sigma_{wi}(t)}
  \Couter_1(y,t)\, dy,
\end{align*}
where we have defined
\begin{gather}
  \xi(t) = \frac{1}{\zeta + \sigma_{gw}(t)}.
  \label{eq:xi}
\end{gather}
Using a similar argument to the leading order solution, we obtain
\begin{gather}
  \Couter_1(y,t) = \frac{\partial \Couter_0}{\partial t}(y,t) \left[ \frac{y^2}{2} -
    \sigma_{wi} y - \frac{\frac{\sigma_{gw}^2}{2} - \sigma_{wi}
      \sigma_{gw} + \xi \left( \frac{\sigma_{wi}^3 - \sigma_{gw}^3}{6} -
        \frac{\sigma_{wi}^2 - \sigma_{gw}^2}{2} \right)}{1 + \xi\,
      (\sigma_{wi} - \sigma_{gw})} \right] .
  \label{eq:outer1}
\end{gather}
Note that $\partial_t \Couter_0 < 0$ so that $\Couter_1$ is an
increasing and concave downward function of $y$ that attains its maximum
value at the right-hand endpoint $y=\sigma_{wi}$; therefore, the
asymptotic solution exhibits the same behaviour observed earlier in the
numerical results for concentration in Figure~\ref{fig:41}b.

\subsubsection{Inner expansion (small time)}
For much shorter times with $t=O(\epsilon)$, we rescale the time variable
according to 
\begin{gather}
  \tau = \frac{t}{\epsilon},
  \label{eq:tau}
\end{gather}
and also denote the inner solution for concentration by
$\Cinner(y,\tau)=C(x,t)$, where $y$ is the same rescaled spatial
variable in \eqref{eq:y}.  Under this scaling the concentration
diffusion equation \eqref{adimc:1-2} reduces to
\begin{gather}
  \frac{\partial \Cinner}{\partial\tau} = \frac{\partial^2
    \Cinner}{\partial y^2}.
  \label{adimc:1-3}
\end{gather}
As mentioned before, over such a short time interval the phase
interfaces are essentially stationary so that we can look for a solution
$\Cinner$ on the fixed interval $y\in [\sigma_{gw}(0),\sigma_{wi}(0)] =
[0,1]$.  The initial and boundary conditions
\eqref{eq:Cnondim-ic}--\eqref{eq:Cnondim-bc} may then be written in
terms of $\Cinner$ as
\begin{align*}
  \Cinner(y,0) &= \Cinit(y),\\
  \Cinner(0,\tau) &= H + \frac{H}{\zeta} \left( \int_0^1 \Cinit(y)\, dy
    - \int_0^1 \Cinner(y,\tau)\, dy \right),\\
  \frac{\partial \Cinner}{\partial y}(1,\tau) &= 0.
\end{align*}

We begin by determining the steady state solution for this problem,
which is simply the constant value
\begin{gather*}
  \Cinner_\infty = \frac{H\zeta + H \ds{\int_0^1 \Cinit(y)\, dy}}{\zeta
    + H}. 
\end{gather*}
We then define $\hat{\Cinner}(y,\tau) = \Cinner(y,\tau) -
\Cinner_\infty$, which satisfies the same equation \eqref{adimc:1-3}
along with the following modified initial and boundary conditions
\begin{align*}
  \hat{\Cinner}(y,0) &= \Cinit(y) - \Cinner_\infty,\\
  \hat{\Cinner}(0,\tau) &= -\frac{H}{\zeta} \int_0^1 \hat{\Cinner}(y,\tau)\, dy,\\
  \frac{\partial \hat{\Cinner}}{\partial y}(1,\tau) &= 0.
\end{align*}
This modified problem can be solved by the method of separation of
variables to obtain 
\begin{gather}
  \hat{\Cinner}(y,\tau) = \sum_{n=1}^{\infty} a_n
  \cos{\big(\mu_n(y-1)\big)} e^{-\mu_n^2\tau}, 
  \label{eq:yhat-series}
\end{gather}
where $\mu_n$ are solutions to the nonlinear equation
\begin{gather}
  \mu_n\zeta + H\tan\mu_n = 0.
  \label{eq:mun}
\end{gather}

In the method of separation of variables, it is customary to determine
the series coefficients $a_n$ by multiplying the initial condition
\begin{gather*}
  \Cinit(y) - \Cinner_\infty = \sum_{n=1}^{\infty}
  a_n\cos{\big(\mu_n(y-1)\big)}  
\end{gather*}
by another eigenfunction from the set $\mathcal{F} =
\{\cos(\mu_n(y-1))\;|\;n=1,2, \dots\}$, then integrating and applying an
orthogonality relation to simplify the result.  We note that
$\mathcal{F}$ is an orthonormal set of eigenfunctions for the diffusion
problem with mixed (Dirichlet/Neumann) and \emph{homogeneous} boundary
conditions, where the eigenvalues are $\mu_n=(2n-1)\frac{\pi}{2}$.  In
contrast, the eigenfunctions in the problem at hand are not orthogonal
because of the integral boundary condition at $y=0$ that leads to the
more complicated eigenvalue equation \eqref{eq:mun} for which the
$\mu_n$ only approach $(2n-1)\frac{\pi}{2}$ as $n\rightarrow \infty$.
As a result, the eigenfunctions satisfy 
\begin{gather*}
  \int_0^1\cos(\mu_n(y-1))\cos(\mu_\ell(y-1))\, dy =
  \begin{cases}
    \frac{1}{2}+\frac{\zeta}{2}\cos^2(\mu_n), & \text{if $n=\ell$},\\ 
    \zeta \cos(\mu_n)\cos(\mu_\ell),          & \text{if $n\neq\ell$}.
  \end{cases}
\end{gather*}
If the eigenfunctions were orthogonal, then the integrals for these two
cases would instead evaluate to $\frac{1}{2}$ and 0 respectively.
For the specific case with $n=\ell=1$, we find that 
\begin{gather*}
  \int_0^1\cos^2(\mu_1(y-1))\, dy \approx 0.4994,
\end{gather*}
while for $n=1$ and $\ell=2$ 
\begin{gather*}
  \int_0^1\cos(\mu_1(y-1))\cos(\mu_2(y-1))\, dy \approx 3.701\times
  10^{-4}. 
\end{gather*}
For larger values of $n$ and $\ell$, these integrals are even closer to
the ideal values of $\frac{1}{2}$ and 0 and therefore the eigenfunctions
are very nearly orthogonal.  As a result, we are able in practice to
evaluate the series coefficients numerically by assuming that they are
orthogonal and taking the inner solution to be
\begin{gather}\label{inner}
  \Cinner(y,\tau) = \frac{H\zeta + H \int_{0}^1 \Cinit(y)\,dy}{\zeta + H}
  + \sum_{n=1}^{\infty} a_n\cos{\big(\mu_n(y-1)\big)}
  e^{-\mu_n^2\tau},   
\end{gather}
where
\begin{gather}
  a_n \approx 2 \int_0^1 (\Cinit(y)-\Cinner_\infty)
  \cos{\big(\mu_n(y-1)\big)}\, dy,
  \label{eq:an}
\end{gather}
and $\mu_n$ are the roots of \eqref{eq:mun}.

We remark here that a similar problem with an integral boundary
condition has been studied by Beilin~\cite{beilin-2001}, who also looked
for a series solution and obtained eigenfunctions that are not
orthogonal.  However, he carried the analytical solution further by
deriving a second set of dual eigenfunctions for an associated adjoint
problem that are orthogonal to the original eigenfunctions.  He then
used both sets of eigenfunctions to calculate the series coefficients
analytically.  We have not applied Beilin's approach here because our
problem has a more complicated integral boundary condition that leads to
a time-dependent boundary condition in the adjoint problem for which we
cannot obtain the eigenfunctions in the same way.

Finally, we note that contrary to the usual approach for developing
matched asymptotics expansions, the inner and outer solutions in our
situation involve no unspecified constant(s) that require matching.  In
particular, the inner solution for the gas concentration tends to the
constant function $\Cinner_\infty$ as $t\rightarrow \infty$.  This is
also the steady state solution of the diffusion equation in the domain
$0\leq y\leq 1$ which coincides with the zeroth order term in the outer
expansion as $t\rightarrow 0$.

\subsection{Comparison with numerical simulations}

\begin{figure}
  \footnotesize
  \centering
  \begin{tabular}{cc}
    (a) Temperature & 
    (b) Gas concentration (short time) \\
    (crosses -- analytical solution). & 
    (crosses -- analytical solution).\\
    \includegraphics[width=0.45\textwidth]{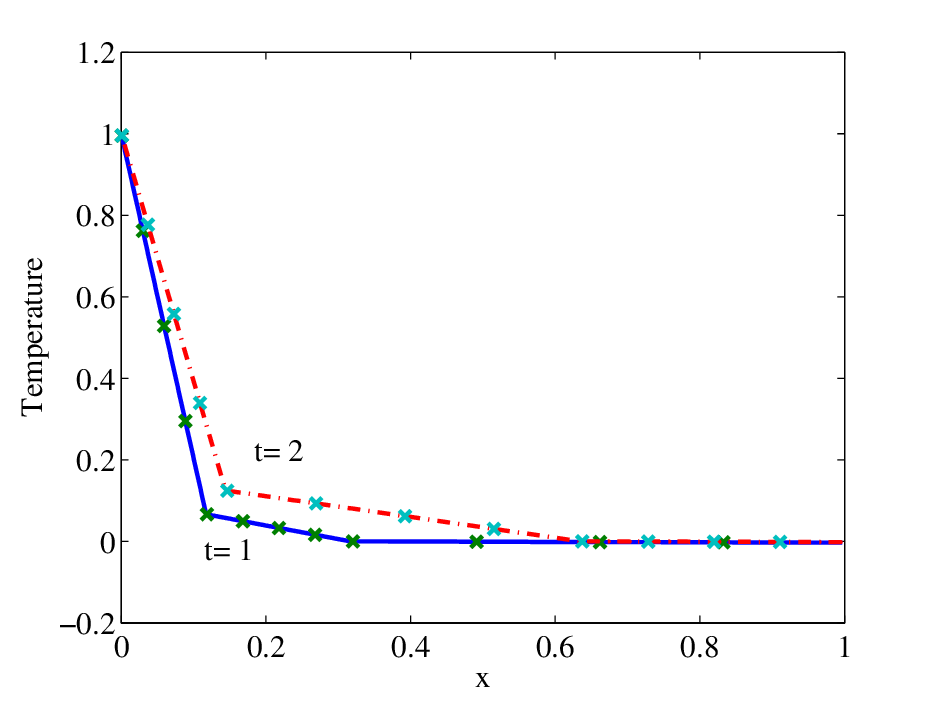}
    & 
    \includegraphics[width=0.45\textwidth]{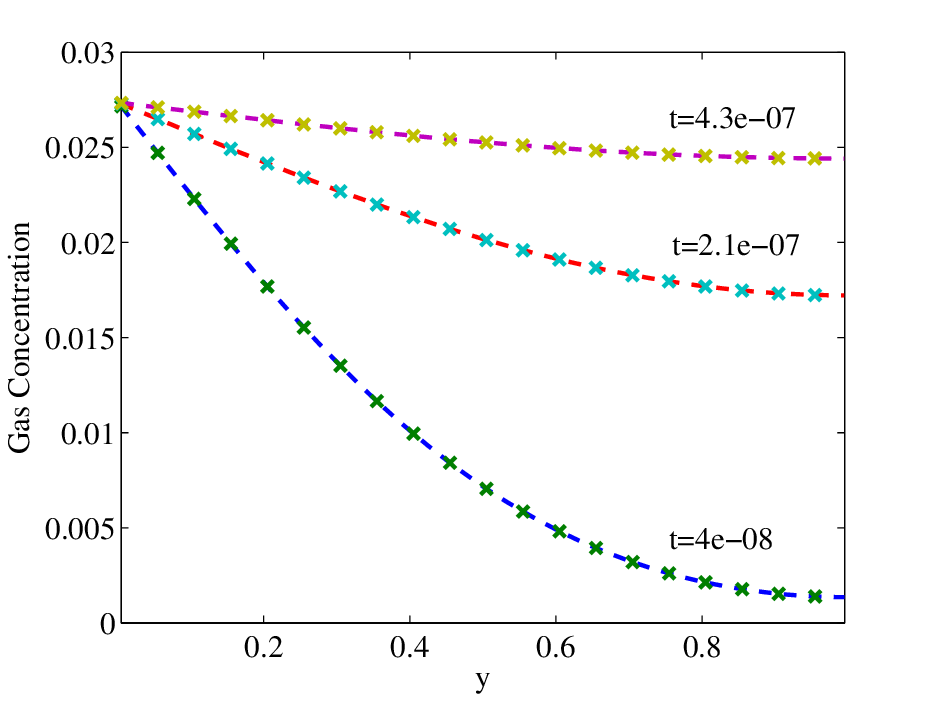}\\
    (c) Water-ice interface. & 
    (d) Gas concentration at $x=L$. \\
    \includegraphics[width=0.45\textwidth]{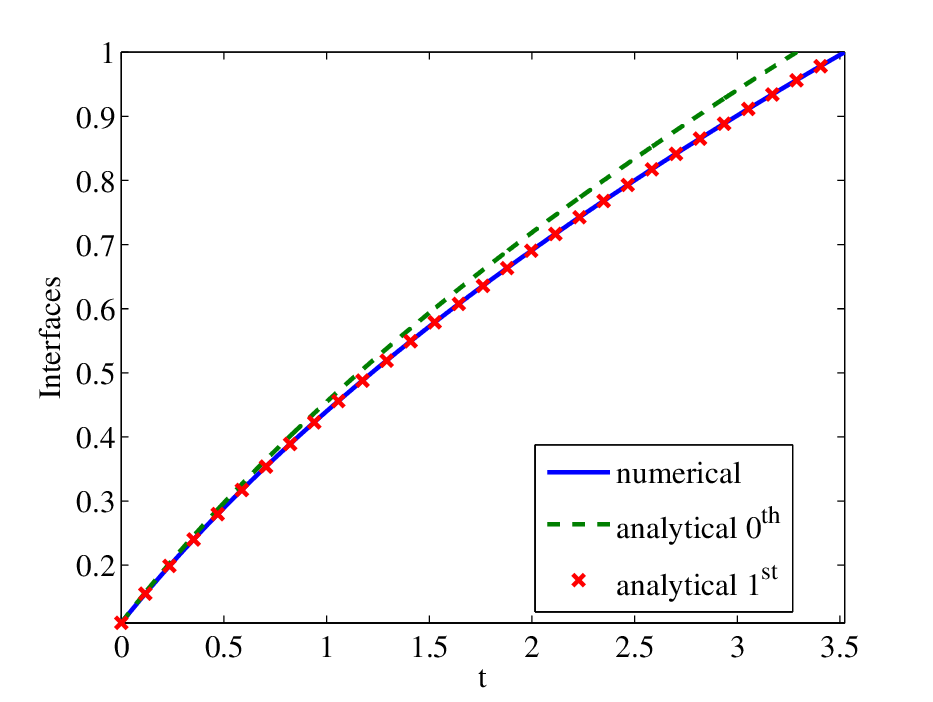}
    & 
    \includegraphics[width=0.45\textwidth]{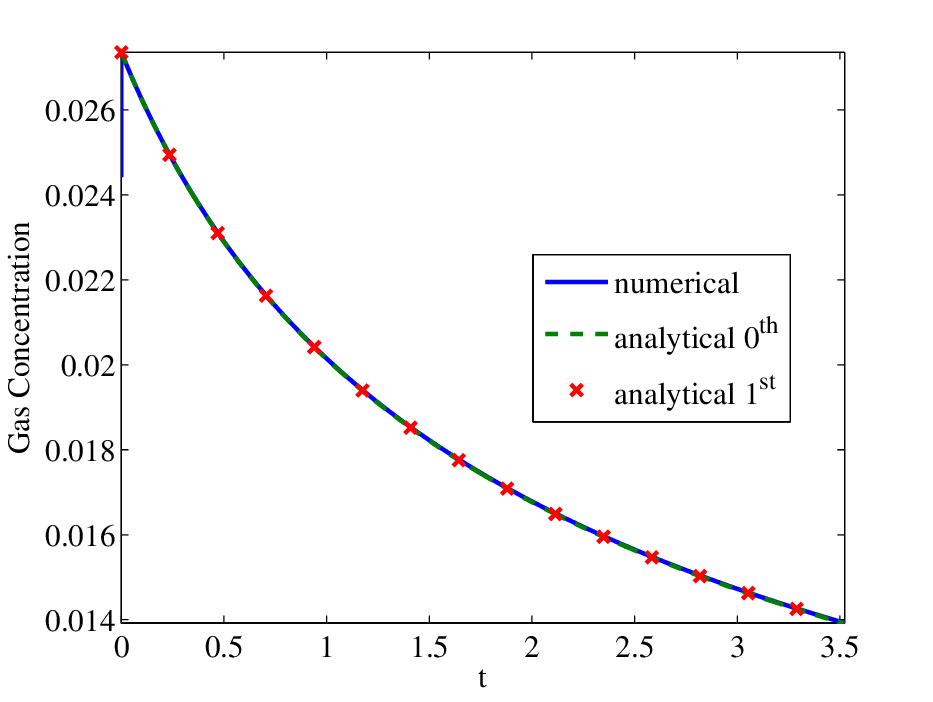}
  \end{tabular}
  \caption{Comparison of analytical and numerical solutions with
    boundary temperatures $T_1=T_c+0.005$, $T_2=T_c-0.005$,
    $\tilde{T}_2=-1$.}
  \label{fig:5}
\end{figure}

\begin{figure}
  \footnotesize
  \centering
  \begin{tabular}{cc}
    (a) Water-ice interface. & (b) Gas concentration at $x=L$. \\
    \includegraphics[width=0.45\textwidth]{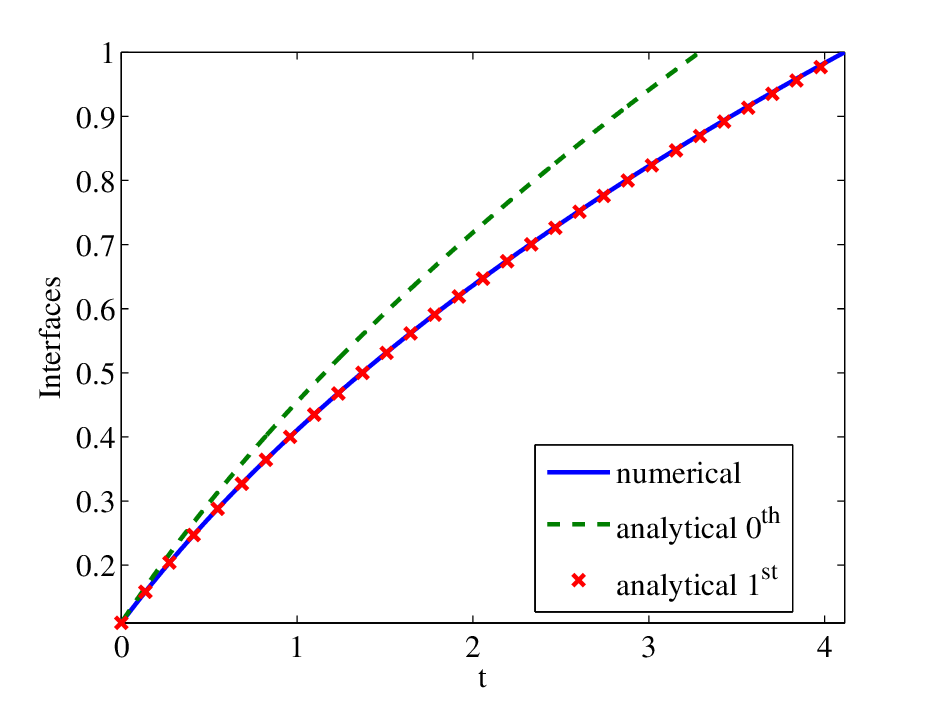}
    & 
    \includegraphics[width=0.45\textwidth]{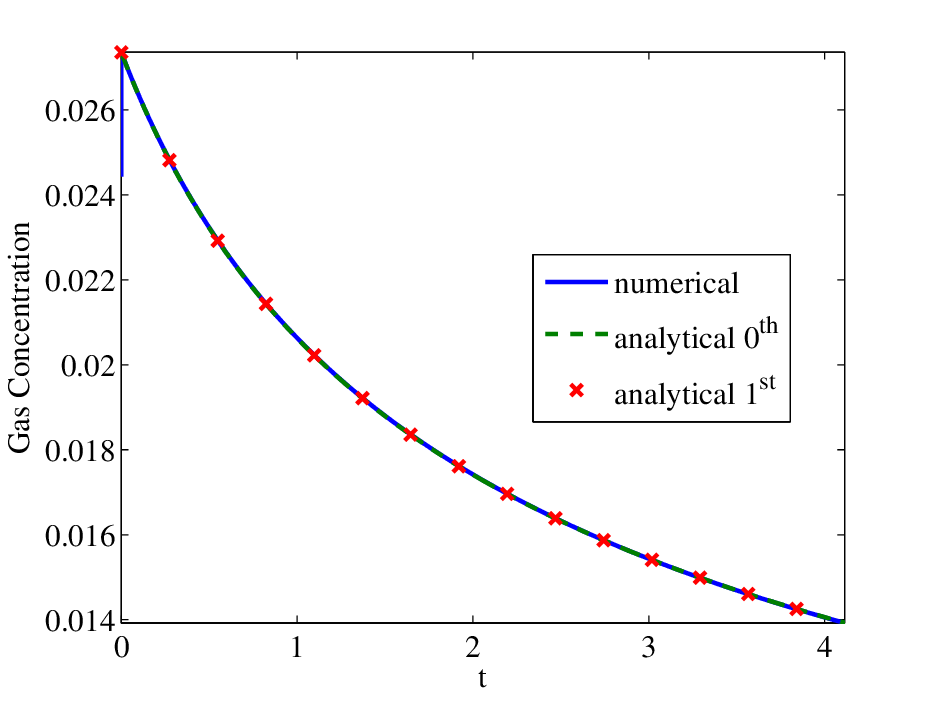}
  \end{tabular} 
  \caption{Comparison of analytical and numerical solutions with
    boundary temperatures $T_1=T_c+0.005$, $T_2=T_c-0.02$,
    $\tilde{T}_2=-4$.}
  \label{fig:6}
\end{figure}

The asymptotic solution developed in the preceding sections is now
calculated using the same parameter values that were used in the full
numerical simulations shown in Figures~\ref{fig:2}--\ref{fig:41}, and
the corresponding results are reported in
Figures~\ref{fig:5}--\ref{fig:8} respectively.  In all cases, the inner
series solution from \eqref{inner} was truncated at 10 terms, while the
outer solution is depicted for both the one- and two-term series
approximations.

\leavethisout{
\begin{figure}
  \footnotesize
  \centering
  \begin{tabular}{cc}
    (a) Fast time scale of gas concentration in water & 
    (b) Slow time scale of gas concentration in water \\
    (crosses -- analytical solutions). & at $x=L$.\\
    \includegraphics[width=0.45\textwidth]{epsfiles/bubble3_21082012_Tinit_T2a-05_Conc_fast_comp.eps}
    & 
    \includegraphics[width=0.45\textwidth]{epsfiles/bubble3_21082012_Tinit_T2a-05_Conc_slow_comp.eps}\\
    (c) Water-ice interface & 
    (d) Temperature (crosses -- analytical solutions) \\
    \includegraphics[width=0.45\textwidth]{epsfiles/bubble3_21082012_Tinit_T2a-05_Interfaces_comp.eps}
    & 
    \includegraphics[width=0.45\textwidth]{epsfiles/bubble3_21082012_Tinit_T2a-05_Temp_comp.eps}
  \end{tabular}
  \caption{Comparison of analytical and numerical solutions with
    boundary temperatures $T_1=T_c+0.01$, $T_2=T_c-0.005$,
    $\tilde{T}_2=-0.5$.}
  \label{fig:7}
\end{figure} 
} 

Focusing first on the base case results in Figure~\ref{fig:5}b, for very
short times the inner concentration solution is indistinguishable to the
naked eye from the computed results.  Over longer times, the temperature
and two-term series expansions for both interfacial position and
concentration (in Figures~\ref{fig:5}a, c and d respectively) also sit
directly on top of the computed results.  The leading order
concentration solution begins to deviate from the computed results when
the boundary temperature difference is increased to $T_2-T_1=-2$ in
Figure~\ref{fig:8}b; this reduction in accuracy derives from the fact
that the $\Couter_0$ approximation is constant in space, whereas the
actual concentration becomes more concave in $y$ as $T_2-T_1$ increases.
There is a more noticeable error in the leading order term for the
interface position, which most evident in Figure~\ref{fig:6}a.

\begin{figure}
  \footnotesize
  \centering
  \begin{tabular}{cc}
    (a) Water-ice interface. & (b) Gas concentration at $x=L$. \\
    \includegraphics[width=0.45\textwidth]{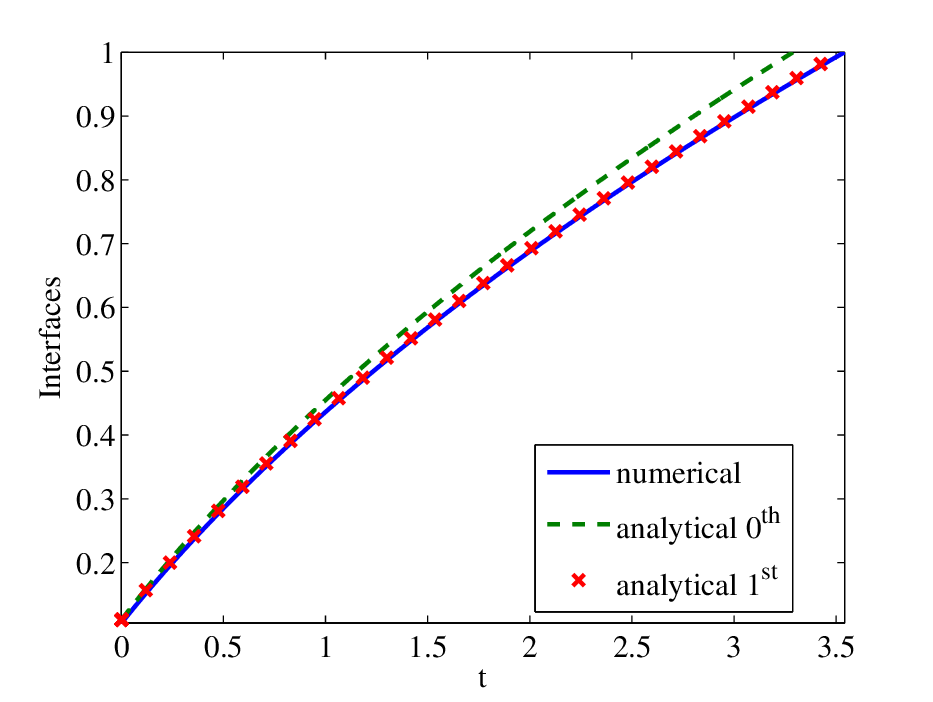}
    &
    \includegraphics[width=0.45\textwidth]{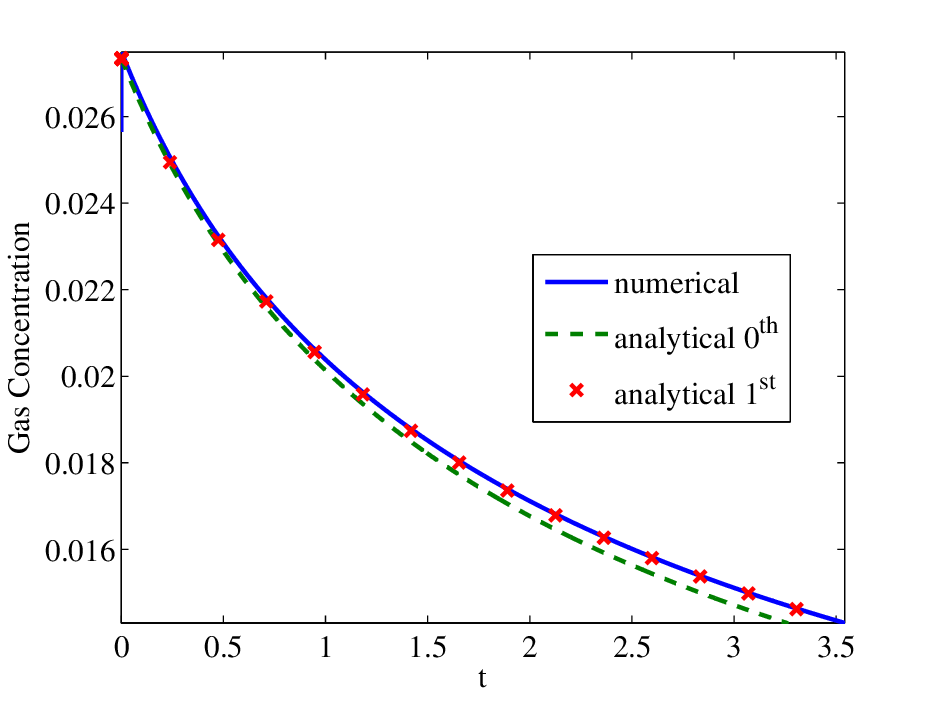}
  \end{tabular}
  \caption{Comparison of analytical and numerical solutions with
    boundary temperatures $T_1=T_c+1$, $T_2=T_c-1$,
    $\tilde{T}_2=-1$.}   
  \label{fig:8}
\end{figure}
 
\begin{figure}
  \footnotesize
  \centering
  \begin{tabular}{cc}
    (a) Water-ice interface. & (b) Gas concentration at $x=L$. \\
    \includegraphics[width=0.45\textwidth]{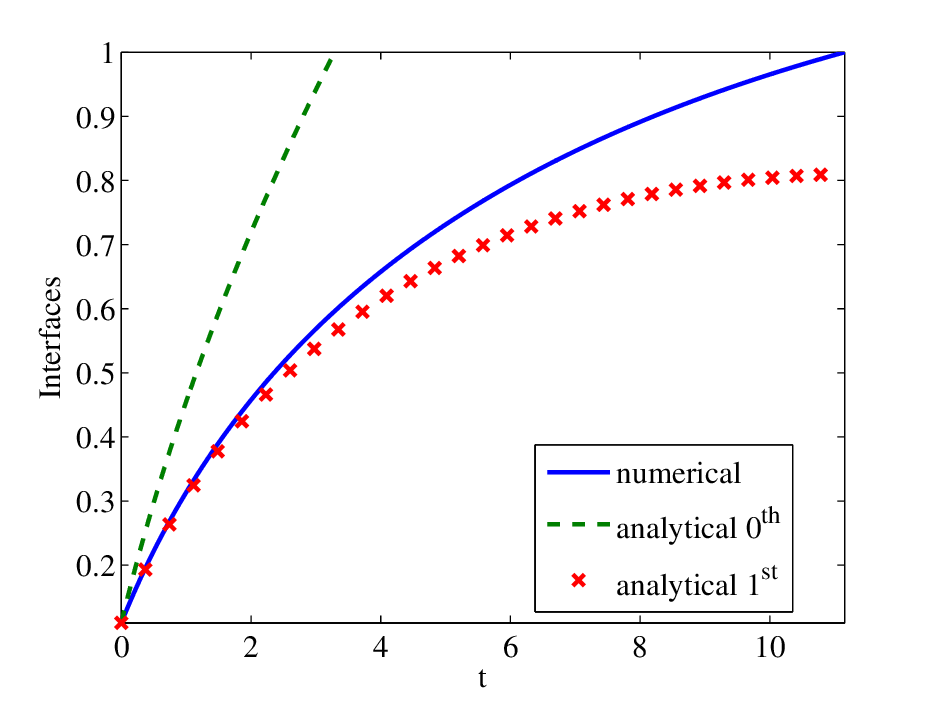}
    &
    \includegraphics[width=0.45\textwidth]{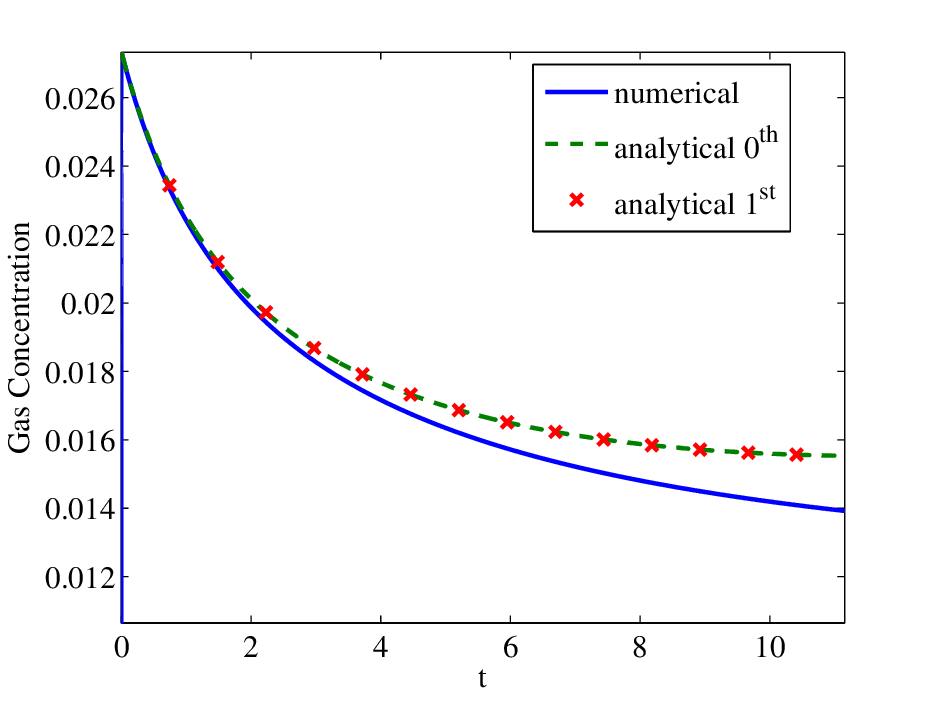}
  \end{tabular}
  \caption{Comparison of analytical and numerical solutions with
    boundary temperatures $T_1=T_c+0.005$, $T_2=T_c-0.005$,
    $\tilde{T}_2=-1$, and $L=1\;cm$.}  
  \label{fig:9}
\end{figure}

It is interesting to investigate the limitations of our asymptotic
solution for more extreme values of the parameters and thereby determine
under what circumstances the series begins to break down.  For example,
if the domain length is increased by several orders of magnitude to
$L=1\;cm$, then there is finally a noticeable error in the two-term
solution for concentration as shown in Figure~\ref{fig:9}b; furthermore,
the two-term asymptotic solution fails to adequately capture the
interface position.  Because it is only for such extreme values of
parameters that the series approximation breaks down, we conclude that
our approximate solutions remain accurate for the range of parameters
corresponding to the melting of frozen sap in maple xylem cells.

\section{Conclusions}
\label{sec:conclusions}

In this paper, we have developed a mathematical model for a three-phase
free boundary problem that is motivated by the study of melting of
frozen sap within maple trees.  The model incorporates both melting of
ice and dissolution of gas within the meltwater.  We derive an
approximate solution that captures the dynamics of the ice-water
interface as a series expansion in the Biot number.  The dissolved gas
concentration exhibits variations over two widely disparate time scales,
leading to a two-scale asymptotic solution.  Comparisons with numerical
simulations show that the approximate solutions are accurate for the
range of parameter values of interest in maple trees.

There are several possible avenues for future work.  First, the
gas-water interfaces within actual xylem cells experience a large
curvature, so that the interfacial surface tension will have a
significant effect on pressure differences.  We would like to include
this effect, as well as the Gibbs-Thompson phenomenon for the variation
of melting temperature across a curved interface which has been
well-studied in the mathematical literature~\cite{luckhaus-1990}.
Secondly, maple trees undergo repeated daily cycles of freezing and
thawing, and so the freezing mechanism also needs to be analysed with a
daily periodic variation in the temperature.  Finally, we would like to
study further some of the technical issues surrounding the extension of
Beilin's approach~\cite{beilin-2001} to the more complicated adjoint
problem that derives from our integral boundary condition.

\appendix
\section{Derivation of the gas-water interface equation \eqref{eq:sgw-dim}}
\label{app:1}

Here we apply a conservation of mass argument to derive the equation
\eqref{eq:sgw-dim} relating $\dot{s}_gw$ and $\dot{s}_{wi}$, assuming
that the domain is a cylinder with constant radius $r$.  At any time
$t$, the total mass of gas is given by the integral
\begin{gather}\label{eq:masgas}
  \mathcal{M}_g(t) 
  = A \left(\int_0^{s_{gw}(t)}\rho_g(s,t)\, ds+m_w(t)\right) 
  = A s_{gw}(0)\rho_g(0),
\end{gather}
while that for water is
\begin{gather}\label{eq:maswat}
  \mathcal{M}_w(t) = A \int_{s_{gw}(t)}^{s_{wi}(t)}\rho_wds 
  = 2A (s_{wi}(t)-s_{gw}(t))\rho_w 
\end{gather}
and for ice is
\begin{gather}\label{eq:masice}
  \mathcal{M}_i(t) = A \int_{s_{wi}(t)}^L\rho_i ds 
  = 2A (L-s_{wi}(t))\rho_i. 
\end{gather}
As mentioned earlier, the water and ice densities are taken to be
constant.  

Since we assume that the system is closed, the sum of the
three masses must be some constant, say $\mathcal{M}_0$, and so
\begin{gather}\label{eq:totalmass}
  \mathcal{M}_0=\mathcal{M}_g(t)+\mathcal{M}_w(t)+\mathcal{M}_i(t).
\end{gather}
Differentiating this expression with respect to time yields
\begin{eqnarray}\label{eq:totalmassder}
  0 &=&0+ A(\dot{s}_{wi}(t)-\dot{s}_{gw}(t))\rho_w - A\dot{s}_{wi}(t)\rho_i, 
\end{eqnarray}
or simply
\begin{gather}
  \dot{s}_{gw}(t)=\left(1-\frac{\rho_i}{\rho_w}\right)\dot{s}_{wi}(t).
\end{gather}

\section{Approximation of the eigenvalues $\mu_n$}

Here we approximate the coefficients $\mu_n$ from equation
\eqref{eq:mun} for small values of $H$.  Then the integral boundary
condition reduces to a pure Dirichlet condition and we then expect that
the eigenvalues and eigenfunctions will reduce to those of the standard
separation of variables solution.  Indeed, if $\frac{H}{\zeta}=0$ then
equation \eqref{eq:mun} reduces to $\mu_n\cos(\mu_n)=0$, whose solutions
are $\{\mu_n^0={(2n-1)}\frac{\pi}{2},\;n=1,2,\dots\}$.
Because we are interested in the case when $\frac{H}{\zeta}$ is very
small, we can make the ansatz $\mu_n=\mu_n^0+\epsilon_n$
with $\epsilon_n\rightarrow 0$, and assume further that
$|\sin(\mu_n)|\approx 1$.  As a result, equation \eqref{eq:mun} becomes
\begin{gather}\label{eq:mun_app}
  |\mu_n| \cdot |\cos(\mu_n)| \approx \frac{H}{\zeta}.
\end{gather}
Using the Taylor series expansion of the cosine function centered at
$\mu_n^0$, \eqref{eq:mun_app} reduces to
\begin{gather}
  \epsilon_n^2 + \mu_n^0\epsilon_n - \frac{H}{\zeta} = 0,
\end{gather}
resulting in 
\begin{gather}
  \epsilon_n =
  \frac{1}{2}\left[\sqrt{(\mu_n^0)^2 + \frac{4H}{\zeta}}-\mu_n^0\right] =
  \frac{\mu_n^0}{2}\left[\sqrt{1+\frac{4H}{\zeta(\mu_n^0)^2}}-1\right].  
\end{gather}
Finally, we employ the approximation
\begin{gather*}
  \sqrt{1+z} = 1+\frac{z}{2} + o(z),
\end{gather*}
to obtain
\begin{gather}\label{eq:error}
  \epsilon_n \approx
  \frac{\mu_n^0}{2}\left[1+\frac{2H}{\zeta(\mu_n^0)^2}-1\right] =
  \frac{H}{\zeta\mu_n^0}, 
\end{gather}
so that
\begin{gather}
  \mu_n = (2n-1)\frac{\pi}{2} + \frac{2 H}{\zeta(2n-1)\pi}.
\end{gather}

\begin{acknowledgements}
  This work was supported by a Discovery Grant from the Natural Sciences
  and Engineering Research Council of Canada and a Research Grant from
  the North American Maple Syrup Council.  MC was funded partially by a
  Fellowship from the Mprime Network of Centres of Excellence.
\end{acknowledgements}

\bibliographystyle{abbrv}
\addcontentsline{toc}{section}{Bibliography}
\bibliography{all}

\begin{thebibliography}{10}

\bibitem{beilin-2001}
S.~A. Beilin.
\newblock Existence of solutions for one-dimensional wave equations with
  nonlocal conditions.
\newblock {\em Electron. J. Diff. Equat.}, 2001(76):1--8, 2001.

\bibitem{Carslaw1988}
H.~S. Carslaw and J.~C. Jaeger.
\newblock {\em Conduction of Heat in Solids}.
\newblock Oxford Science Publications. Clarendon Press, New York, second
  edition, 1988.

\bibitem{Ceseri_Stockie2012}
M.~Ceseri and J.~M. Stockie.
\newblock A mathematical model for sap exudation in maple trees governed by ice
  melting, gas dissolution and osmosis.
\newblock To appear in \textit{SIAM J. Appl. Math.}, 2012.

\bibitem{Crank1956}
J.~Crank.
\newblock {\em The Mathematics of Diffusion}.
\newblock Clarendon Press, 1956.

\bibitem{Crank1984}
J.~Crank.
\newblock {\em Free and Moving Boundary Problems}.
\newblock Clarendon Press, New York, 1984.

\bibitem{Friedman1959}
A.~Friedman.
\newblock Free boundary problems for parabolic equations. {I}. {M}elting of
  solids.
\newblock {\em J. Math. Mech.}, 8:499--517, 1959.

\bibitem{Friedman1960}
A.~Friedman.
\newblock Free boundary problems for parabolic equations. {III}. {D}issolution
  of a gas bubble in liquid.
\newblock {\em J. Math. Mech.}, 9:327--345, 1960.

\bibitem{friedman-1982}
A.~Friedman.
\newblock {\em Variational Principles and Free-Boundary Problems}.
\newblock John Wiley \&\ Sons, New York, 1982.

\bibitem{friedman-2000}
A.~Friedman.
\newblock Free boundary problems in science and technology.
\newblock {\em AMS Notices}, 47(8):854--861, 2000.

\bibitem{MR605400}
R.~M. Furzeland.
\newblock A comparative study of numerical methods for moving boundary
  problems.
\newblock {\em J. Inst. Math. Appl.}, 26(4):411--429, 1980.

\bibitem{gupta-2003}
S.~C. Gupta.
\newblock {\em The Classican Stefan Problem: Basic Concepts, Modelling and
  Analysis}, volume~45 of {\em North-Holland Series in Applied Mathematics and
  Mechanics}.
\newblock Elsevier, Amsterdam, 2003.

\bibitem{MR2722625}
W.~Huang and R.~D. Russell.
\newblock {\em Adaptive Moving Mesh Methods}, volume 174 of {\em Applied
  Mathematical Sciences}.
\newblock Springer, New York, 2011.

\bibitem{Huyakorn1994}
P.~Huyakorn, S.~Panday, and Y.~Wu.
\newblock A three-dimensional multiphase flow model for assesing {NAPL}
  contamination in porous and fractured media, 1. {F}ormulation.
\newblock {\em J. Contam. Hydrol.}, 16(2):109--130, 1994.

\bibitem{Keller1964}
J.~B. Keller.
\newblock Growth and decay of gas bubbles in liquids.
\newblock In {\em Proceedings of the Symposium on Cavitation in Real Liquids
  (General Motors Research Laboratory, Warren, MI)}, pages 19--29. Elsevier,
  New York, 1964.

\bibitem{konrad-rothnebelsick-2003}
W.~Konrad and A.~Roth-Nebelsick.
\newblock The dynamics of gas bubbles in conduits of vascular plants and
  implications for embolism repair.
\newblock {\em J. Theor. Bio.}, 224:43--61, 2003.

\bibitem{luckhaus-1990}
S.~Luckhaus.
\newblock Solutions for the two-phase {S}tefan problem with the
  {G}ibbs-{T}homson {L}aw for the melting temperature.
\newblock {\em Euro. J. Appl. Math.}, 1:101--111, 1990.

\bibitem{Milburn1984}
J.~Milburn and P.~O'Malley.
\newblock Freeze-induced fluctuations in xylem sap pressure in \textit{Acer
  pseudoplatanus}: {A} possible mechanism.
\newblock {\em Can. J. Bot.}, 62:2100--2106, 1984.

\bibitem{plesset-prosperetti-1977}
M.~S. Plesset and A.~Prosperetti.
\newblock Bubble dynamics and cavitation.
\newblock {\em Annu. Rev. Fluid Mech.}, 9:145--185, 1977.

\bibitem{Singh2012}
K.~Singh and R.~K. Niven.
\newblock Non-aqueous phase liquid spills in freezing and thawing soils:
  {C}ritical analysis of pore-scale processes.
\newblock {\em Crit. Rev. Environ. Sci. Technol.}, 2012.
\newblock In press, DOI: 10.1080/10643389.2011.604264.

\bibitem{Tao1979}
L.~N. Tao.
\newblock On solidification problems including the density jump at the moving
  boundary.
\newblock {\em Quart. J. Mech. Appl. Math.}, 32(2):175--185, 1979.

\bibitem{tsypkin2000}
G.~G. Tsypkin.
\newblock Mathematical models of gas hydrates dissociation in porous media.
\newblock {\em Ann. New York Acad. Sci.}, 912(1):428--436, 2000.

\bibitem{tyree-sperry-1989}
M.~T. Tyree and J.~S. Sperry.
\newblock Vulnerability of xylem to cavitation and embolism.
\newblock {\em Annu. Rev. Plant Physiol. Plant Mol. Biol.}, 40:19--38, 1989.

\bibitem{Wilson1982}
D.~G. Wilson.
\newblock One dimensional multi-phase moving boundary problems with phases of
  different densities.
\newblock Technical Report CSD-93, Oak Ridge National Laboratory, January 1982.

\bibitem{xu-2004}
W.~Xu.
\newblock Modeling dynamic marine gas hydrate systems.
\newblock {\em Amer. Mineral.}, 89(8--9):1271--1279, 2004.

\end{thebibliography}

\end{document}